\documentclass[]{aa} 

\usepackage[utf8]{inputenx}
\usepackage{graphicx}
\usepackage{txfonts}

\usepackage{color}
\usepackage{url}
\usepackage{arydshln}
\usepackage{grffile}

\begin{document} 

\title{The MUSE-Wide Survey: A determination of the Lyman $\alpha$
  emitter luminosity function at $3 < z < 6$\thanks{Based on
    observations collected at the European Organisation for
    Astronomical Research in the Southern Hemisphere under ESO
    programme 094.A-025.}}

\titlerunning{MUSE-Wide:  Lyman $\alpha$ Luminosity Function}
\authorrunning{E.~C.~Herenz et al.}

\author{
  Edmund~Christian~Herenz\inst{\ref{inst1}} \and
  Lutz~Wisotzki\inst{\ref{inst2}} \and
  Rikke~Saust\inst{\ref{inst2}} \and
  Josephine~Kerutt\inst{\ref{inst2}} \and
  Tanya~Urrutia\inst{\ref{inst2}} \and
  Catrina Diener\inst{\ref{inst3}} \and
  Kasper~Borello~Schmidt\inst{\ref{inst2}} \and
  Raffaella~Anna~Marino\inst{\ref{eth}} \and
  Geoffroy~de~la~Vieuville\inst{\ref{irap}} \and
  Leindert Boogaard\inst{\ref{inst5}} \and
  Joop~Schaye\inst{\ref{inst5}} \and
  Bruno~Guiderdoni\inst{\ref{inst4}} \and
  Johan~Richard\inst{\ref{inst4}} \and
  Roland~Bacon\inst{\ref{inst4}}
}
\institute{
  Department of Astronomy, Stockholm University,
  AlbaNova University Centre, SE-106 91,
  Stockholm, Sweden
  \label{inst1}
  \and
  Leibniz-Institut für Astrophysik Potsdam (AIP), An der
  Sternware 16, 14482 Potsdam,
  Germany
  \label{inst2}
  \and
  University of Cambridge, Madingley Road, Cambridge CB3 0HA,
  UK
  \label{inst3}
  \and
  ETH Z\"{u}rich, Department of Physics, Wolfgang-Pauli-Strasse 27,
  8093 Z\"{u}rich, Switzerland
  \label{eth}
  \and
  Institut de Recherche en Astrophysique et Plan\'{e}tologie (IRAP),
  Universit\'{e} de Toulouse, CNRS, UP, CNES, 14 avenue Edouard Belin,
  F-31400 Toulouse, France
  \label{irap}
  \and
  Leiden Observatory, Leiden University, PO Box 9513, 2300 RA Leiden,
  The Netherlands
  \label{inst5}
  \and
  Univ Lyon, Univ Lyon1, Ens de Lyon, CNRS, Centre de Recherche
  Astrophysique de Lyon UMR5574, F-69230, Saint-Genis-Laval, France
  \label{inst4}
}

\abstract{We investigate the Lyman $\alpha$ emitter (LAE) luminosity
  function (LF) within the redshift range $2.9 \leq z \leq 6$
  from the first instalment of the blind integral field spectroscopic
   MUSE-Wide survey.  This initial part of the survey probes a region
  of 22.2\,arcmin$^2$ in the CANDELS/GOODS-S field (24 MUSE pointings
  with 1h integrations).  The dataset provided us with 237 LAEs from
  which we construct the LAE LF in the luminosity range
  $42.2 \leq \log L_\mathrm{Ly\alpha} [\mathrm{erg\,s}^{-1}]\leq 43.5$
  within a volume of $2.3\times10^5$\,Mpc$^3$. For the LF construction
  we utilise three different non-parametric estimators: the classical
  $1/V_\mathrm{max}$ method, the $C^{-}$ method, and an improved
  binned estimator for the differential LF.  All three methods deliver
  consistent results, with the cumulative LAE LF being
  $\Phi(\log L_\mathrm{Ly\alpha} [\mathrm{erg\,s}^{-1}] = 43.5) \simeq
  3\times 10^{-6}$\,Mpc$^{-3}$ and
  $\Phi(\log L_\mathrm{Ly\alpha} [\mathrm{erg\,s}^{-1}] = 42.2) \simeq
  2 \times 10^{-3}$\,Mpc$^{-3}$ towards the bright and faint end of
  our survey, respectively.  By employing a non-parametric statistical
  test, and by comparing the full sample to subsamples  in
  redshift bins, we find no supporting evidence for an evolving LAE LF
  over the probed redshift and luminosity range.  Using a parametric
  maximum-likelihood technique we determine the  best-fitting
  Schechter function parameters $\alpha = -1.84^{+0.42}_{-0.41}$ and
  $\log L^* [\mathrm{erg\,s}^{-1}] = 42.2^{+0.22}_{-0.16}$ with the
  corresponding normalisation
  $\log \phi^* [\mathrm{Mpc}^{-3}] = -2.71$.  However, the dynamic
  range in Ly$\alpha$ luminosities probed by MUSE-Wide leads to a
  strong degeneracy between $\alpha$ and $L^*$.  Moreover, we find
  that a power-law parameterisation of the LF appears to be less
  consistent with the data compared to the Schechter function, even so
  when not excluding the X-Ray identified AGN from the sample.  When
  correcting for completeness in the LAE LF determinations, we take
  into account that LAEs exhibit diffuse extended low
  surface brightness halos.  We compare the resulting LF to one
  obtained by applying a correction assuming compact point-like
  emission. We find that the standard correction underestimates the
  LAE LF at the faint end of our survey by a factor of
  2.5. Contrasting our results to the literature we find that at
  $\log L_\mathrm{Ly\alpha} [\mathrm{erg\,s}^{-1}] \lesssim 42.5$
  previous LAE LF determinations from narrow-band surveys appear to be
  affected by a similar bias.  }

\keywords{Cosmology: observations -- Galaxies: high-redshift --
  Galaxies: luminosity function -- Techniques: imaging spectroscopy}

\maketitle

\defcitealias{Herenz2017a}{H017}


\section{Introduction}
\label{sec:introduction}

One of the most fundamental statistical distribution functions to
characterise the population of galaxies is the galaxy luminosity
function (LF).  The differential LF, $\psi(L,z)$,
counts the number of galaxies $N$ per unit volume $V$ as a function of
luminosity $L$ and redshift $z$:
$\mathrm{d}N = \psi(L,z)\,\mathrm{d}L\,\mathrm{d}V\,\mathrm{d}z$.
While this bi-modal form provides the most general description,
observationally the LF is often determined at a fixed redshift or a
redshift interval over which effects of redshift evolution are deemed
negligible, i.e.
\begin{equation}
  \mathrm{d}N = \phi(L)\,\mathrm{d}L\,\mathrm{d}V\;\mathrm{.}
  \label{eq:1}
\end{equation}
Galaxy LFs and their redshift evolution provide a gold standard for
summarising the changing demographics of galaxies with cosmic
look-back time.  Being essential benchmarks for cosmological models of
galaxy formation and evolution in our universe, LF determinations are
often amongst the pivotal goals in the design and analysis of
extragalactic surveys
\citep[][]{Petrosian1992,Willmer1997,Johnston2011,Dunlop2013,Caditz2016}.

While Lyman $\alpha$ (Ly$\alpha$, $\lambda 1216$) emission was already
suggested as a prime tracer for galaxy formation studies in the early
universe more than five decades ago \citep{Partridge1967}, initial
searches for such high-$z$ Ly$\alpha$ emitting galaxies (LAEs) were
unsuccessful and, hence, constrained only upper limits of the
LAE LF \citep[see review by][]{Pritchet1994}.

The first successful detections of LAEs accompanied by spectroscopic
confirmations on 8\,m class telescopes employed a colour-excess
criterion in narrow-band (NB) images \citep{Hu1996,Hu1998}.  In the
following years the NB imaging technique was routinely used to
construct LAE samples of up to several hundreds of galaxies at
$2 \lesssim z \lesssim 5$ \citep[see review by][]{Taniguchi2003}.
Mostly from such NB surveys, sometimes in combination with
spectroscopic follow-up of subsamples, the first LAE LF estimates up
to $z\sim6$ were obtained
\citep[e.g.][]{Cowie1998,Kudritzki2000,Ouchi2003,Hu2004,Ajiki2004,Tapken2006,Dawson2007,Gronwall2007,Murayama2007,Sawicki2008,Henry2012}.

Most of the LAE LF studies so far focused on a single
redshift slice of typically $\Delta z \simeq 0.1$ (corresponding to
typical NB filter widths $\Delta \lambda \simeq 100$\AA{}).
Significant progress in terms of methodology, numbers of LAEs, and
rate of spectroscopic follow-up observations was achieved by
\cite{Ouchi2008} within three redshift slices ($z\sim\{3,4,5\}$) over
a 1\,deg$^2$ region in the \emph{Subaru}/XMM-Newton Deep Survey
\citep[SXDS;][]{Furusawa2008}.  Later, \cite{Ouchi2010} extended the
SXDS LAE survey to $z\approx6.6$.  More recently, further
\emph{Subaru}/Suprime-Cam NB imaging data in other fields were used to
construct LAE LFs over 5\,deg$^2$ at $z=5.7$ and $z=6.6$
\citep{Matthee2015,Santos2016}.  Moreover, by combining NB and
medium-band observations from the Subaru and the Isaac Newton
Telescope \cite{Sobral2017} constructed a LAE LF from $\sim4000$ LAEs
simultaneously from redshifts $z\sim2$ to $z\sim6$.

The latest development in NB LAE surveys is due to the advent of
\emph{Subaru}/Hyper Suprime-Cam, a 1.5\,deg$^{2}$ wide-field imager
\citep{Miyazaki2012,Miyazaki2018}.  Recently, the first
  results for a $\sim$14 deg$^{2}$ and $\sim$21 deg$^{2}$ NB survey at
  $z\sim 6$ and $z\sim 7$, respectively, where published \citep[the so
  called SILVERUSH survey][]{Ouchi2018,Shibuya2018a,Shibuya2018}.
  From this unprecedented dataset \cite{Konno2018} constructed the LAE
  LF for sources $L_\mathrm{Ly\alpha} \gtrsim 10^{43}$erg\,s$^{-1}$.
Without any doubt NB imaging studies have been and are still of
central importance for our understanding of the LAE LF.  Only their
wide nature allows the construction of statistical samples of the
brightest and rarest LAEs.

However, the LAE LF determination from NB imaging studies is fraught
with some difficulties that can be alleviated in blind surveys with an
integral field spectrograph \citep[IFS, see e.g.  the recent textbook
by][]{Bacon2017a}. Especially, since an IFS samples spectra over a
contiguous field of view, the resulting 3D datacubes can be envisioned
as a stack of NB images of much smaller bandwidth compared to
imaging NB filters.  Thus, a blind search for emission line
sources in an IFS datacube directly provides  a catalogue of
spectroscopically confirmed Ly$\alpha$ emitters, avoiding the need for
follow-up spectroscopy.  Then, flux measurements on the lines can be
performed in virtually any conceivable aperture, resulting in reliable
flux measurements absent of slit or bandpass losses.  Moreover,
instead of probing only a tiny redshift slice, IFSs cover an extended
range of redshifts.  Another advantage is that the narrow bandwidth of
the individual wavelength slices in the datacube significantly reduces
the contribution of sky background to emission line signals.  This
allows IFS searches to reach significant fainter limiting fluxes
compared to NB imaging surveys.  Lastly, by construction an integral
field spectroscopic survey delivers a flux-limited LAE sample, rather
than an equivalent-width limited sample.  This mitigates possible
biases from heterogeneous equivalent width cuts in NB imaging studies.

A pilot IFS survey for LAEs between $3 < z < 6$ was performed by
\cite{vanBreukelen2005} with the Visible Multi Object Spectrograph
\citep{LeFevre2003} Integral Field Unit at ESOs Very Large Telescope
(VLT).  However, this pilot study was severely limited by the
relatively low throughput, small field of view, and the low spectral
resolution of this instrument.  Substantial progress in performing a
blind IFS survey to detect Ly$\alpha$ emitters was achieved in the
Hobby Eberle Telescope Dark Energy Experiment (HETDEX) Pilot Survey by
\cite{Adams2011}.  Utilising 61 nights of observations with VIRUS-P
\citep{Hill2008}, a path-finder fibre-fed IFS that will be replicated 156
times for the final HETDEX survey \citep{Hill2016} on the McDonald
2.7m Harlan J. Smith telescope, \citeauthor{Adams2011} constructed a
catalogue of 397 emission line galaxies blindly selected over 169
arcmin$^2$ in areas with rich complementary datasets.  This catalogue
contained 99 LAEs without X-ray counterparts in the range $1.9 < z < 3.8$.
From the \citeauthor{Adams2011} catalogue \cite{Blanc2011} constructed
the Ly$\alpha$ LF in the luminosity range
$42.6 \leq \log L_\mathrm{Ly\alpha} \,[\mathrm{erg\,s}^{-1}] \leq
43.5$.

With the advent of the Multi Unit Spectroscopic Explorer (MUSE) at ESOs VLT
\citep[]{Bacon2014,Caillier2014} the field of blind deep IFS
surveys was revolutionised.  This image-slicer-based IFS with a
1\arcmin{}$\times$1\arcmin{} field of view covering the wavelength
range from 4650\AA{} to 9300\AA{} was designed from the ground up as a
discovery machine for faint emission line galaxies, especially LAEs at
high redshift \citep[$2.9 \lesssim z \lesssim 6.6$,][]{Bacon2004}.

Its unprecedented potential for LAE LF determinations was demonstrated
in the analysis of a 27h integration on the \emph{Hubble} Deep Field
South \citep{Casertano2000} obtained during commissioning
\citep{Bacon2015}.  By utilising 59 LAEs from this dataset
\cite{Drake2017} was able to put constraints on the Ly$\alpha$ LF down to
$\log L_\mathrm{Ly\alpha} [\mathrm{erg\,s}^{-1}] = 41.4$, almost an
order of magnitude deeper than nearly all previous
observational efforts;  the only exception was a heroic 92h
deep long-slit integration with the FORS2 instrument on ESOs VLT
\citep{Rauch2008}.
Recently, the \cite{Drake2017} pilot-study was significantly refined
by \cite{Drake2017a} using 601 LAEs found in the MUSE Consortium
Guaranteed Time Observations \citep[GTO,][]{Bacon2017,Inami2017} of the
\emph{Hubble} Ultra Deep Field \citep{Beckwith2006}.  This dataset
consists of a 3.15\arcmin{}$\times$3.15\arcmin{} mosaic exposed at
10h depth, and a central 1.15\,arcmin$^2$ 31h deep pointing that
reached depths similar to those of the pilot study in the Hubble Deep field
South.  As a novelty \cite{Drake2017a} accounted for the effect of
extended low surface brightness Ly$\alpha$ halos in their
completeness assessment.

However,  the pencil beam nature of the MUSE-deep fields does not allow us
to probe the LAE LF at brighter luminosities.  Thus, complementary to
the MUSE Deep Fields a shallower large-area programme, known as
MUSE-Wide (MW), is also part of the MUSE GTO.  MUSE-Wide  covers
100\,arcmin$^{2}$ at 1h depth in regions where deep HST imaging
surveys were performed, namely the CANDELS/Deep region in the
\emph{Chandra} Deep Field South \citep[CDFS]{Grogin2011,Koekemoer2011}
and the GOODS/South survey \citep{Giavalisco2004}.  Recently,
\cite{Herenz2017a} \citepalias[hereafter][]{Herenz2017a} presented a
catalogue of 831 emission line selected galaxies from the first 24
MUSE-Wide pointings (corresponding to an area of 22.2\,arcmin$^2$) in
the CDFS.  This catalogue contains 237 LAEs with luminosities
$41.6 \leq \log L_\mathrm{Ly\alpha} [\mathrm{erg\,s}^{-1}] \leq 43.5$
in the redshift range $3 < z < 6$, thus augmenting the sample of faint
LAEs from the MUSE-Deep fields.
In the present manuscript we  use the LAE sample obtained in
the first 24 MUSE-Wide pointings to study the LAE LF.

The structure of this manuscript is as follows. In
Sect.~\ref{sec:data-lyman-alpha} we provide an overview of the
 MUSE-Wide survey data used and we describe how we obtained the LAE
initial sample from this dataset.  Following, in
Sect.~\ref{sec:constr-muse-wide} we explain how we constructed the LAE
selection function in MW.  Then, in
Sect.~\ref{sec:lyman-alpha-emitter} we provide an overview of the
 methods adopted for constructing the LAE LF.  Our results on the LAE
LF are given in Sect.~\ref{sec:results}. In
Sect.~\ref{sec:comp-with-liter} we compare our results with those from the
literature.  Finally, we summarise the results obtained so far in
Sect.~\ref{sec:discussion}, where we also present an outlook for
further refinements of our study that will be relevant with the
release of the full MUSE-Wide sample.

Throughout the paper we always assume a standard $\Lambda$CDM
concordance cosmology with $\Omega_\Lambda = 0.7$, $\Omega_M = 0.3$,
and $H_0 = 70$\,km\,s$^{-1}$ when converting observed to physical
quantities.

\section{MUSE-Wide data and Ly$\alpha$ emitter sample}
\label{sec:data-lyman-alpha}

The data under scrutiny in this paper are the 24 adjacent
1\arcmin{}$\times$1\arcmin{} one-hour deep MUSE pointings in the
CANDELS/Deep region of the GOODS-South field.  The data were taken
during the first period of MUSE GTO Observations between September and
October 2014 (ESO programme ID 094.A-0205) as part of the MUSE-Wide
(hereafer MW) survey.  Accounting for the 4\arcsec{} overlap between
individual pointings, the total survey area is 22.2 arcmin$^2$.  The
survey covers a wavelength range from 4750\,\AA{} to 9350\,\AA{}, thus
probing Ly$\alpha$ emitters within the redshift range
$2.9 \leq z \leq 6.7$.

A detailed account of the observations, data reduction, and
construction of the emission line selected galaxy catalogue has been
given in \citetalias{Herenz2017a}; here we only provide an overview.

\subsection{Observations and data reduction}
\label{sec:observ-data-reduct}

Each 1h deep MW pointing consists of four individual 15-minute
exposures.  More than half of the observations were taken under
photometric conditions during dark and grey nights, with the remainder
taken under clear conditions during dark nights.  The measured seeing ranged
from 0.7\arcsec{} to 1.1\arcsec{}, with 0.9\arcsec{} being the average
of the observations.

For each of the pointings a datacube was created by employing the MUSE
data reduction system \citep[][]{Weilbacher2014} in combination with a
few custom developed routines and the \emph{Zurich Atmosphere Purge}
(ZAP) sky-subtraction software\footnote{ZAP is publicly available via
  the Astrophysics Source Code Library: \url{http://ascl.net/1602.003}
  \citep{Soto2016a}.} \citep{Soto2016}.  We also used the
self-calibration procedure that is part of the MUSE Python Data
Analysis Framework -- MPDAF\footnote{MPDAF is publicly available via
  the Astrophysics Source Code Library: \url{http://ascl.net/1611.003}
  \citep{Piqueras2017}.} \citep{Conseil2016,Bacon2017}.

The reduced data consists of 24 datacubes, each covering
1\arcmin{}$\times$1\arcmin{} on the sky with a spatial sampling
0.2\arcsec{}$\times$0.2\arcsec{}.  These spatial sampling elements
(so-called spaxels) contain a spectrum from 4750\,\AA{}--9350\,\AA{} that is sampled
at 1.25\AA{} in air wavelengths.  Each volume element
(called a voxel) of a datacube stores the received flux density in units of
$10^{-20}$erg\,s$^{-1}$cm$^{-2}$\AA{}$^{-1}$.  The full width at half
maximum (FWHM) of the spectrographs line spread function is roughly
twice the spectral sampling (i.e. $\sim$2.5\AA{}) resulting in a
resolving power of $R\sim 1900$--$3800$ over the  wavelength
range covered.

The MUSE data reduction system also propagates the variances during
all reduction steps into each voxel, thereby creating a complementary
variance datacube for each pointing.  However, these formal variance
values underestimate the true variances, and are thus not optimal for
emission line detection and estimation of the error on emission line
flux measurements.  In order to correct for this, we performed an
empirical estimate of the variance values by evaluating the statistics
of randomly placed apertures in empty regions of the sky (see
Sect.~3.1.1 in \citetalias{Herenz2017a}).

\subsection{Emission line galaxy catalogue}
\label{sec:lyman-alpha-emitter-1}

Emission line source detection in the MW data is performed with
our dedicated \emph{Line Source Detection and Cataloguing Tool}
\texttt{LSDCat}\footnote{\texttt{LSDCat} is publicly available via the
  Astrophysics Source Code Library: \url{http://ascl.net/1612.002}
  \citep{Herenz2016a}.} \citep{Herenz2017}.  As a required preparatory
step before emission line source detection we remove source continua
from the datacube by subtracting a $\approx$190\AA{} wide running
median in the spectral direction. This method of removing source
continua has proven to be very effective, leaving as remaining
features mostly real emission lines and straightforward identifiable
residuals from continua that vary at higher frequencies than the width
of the median filter (e.g. cold stars).

In the next step \texttt{LSDCat} cross-correlates each datacube with a
3D matched filter template for compact emission line sources.  We used
a 3D Gaussian as the template, with its spatial FWHM dictated by
the wavelength dependent seeing point spread function (PSF) and its spectral FWHM fixed to
$v_\mathrm{FWHM} = 250$\,km\,s$^{-1}$.  As demonstrated in Sect. 4.3
of \cite{Herenz2017}, this value is optimal for detecting LAEs
in MUSE surveys at their highest possible signal-to-noise ratios (S/N).  Then the initial emission line candidate catalogue was
created by setting the detection threshold to
S/N$_\mathrm{thresh} = 8$.  This initial catalogue was then screened
by four investigators (ECH, LW, TU, and JK) using the interactive
graphical tool \emph{QtClassify}\footnote{\emph{QtClassify} is
  publicly available via the Astrophysics Source Code Library:
  \url{http://ascl.net/1703.011} \citep{Kerutt2017}.}
(\citealt{Kerutt2017}; see also Appendix of \citetalias{Herenz2017a}).
The purpose of this screening process was to identify the detected
emission lines, and to purge obviously spurious detections
(e.g. those due to continuum residuals).  Real detections were assigned with
quality and confidence flags.  Here, the quality flag encodes whether
multiple emission lines of a source were detected (quality A),
multiple emission lines are present but below the detection
threshold (quality B), or whether the identification was based on a
single line (quality C). By this definition all of the LAEs considered
in the present analysis are quality C objects.  As detailed in
\citetalias{Herenz2017a} (Sect.~3.1.4), the confidence values
encode a more subjective measure of belief towards the final
identification of a source, ranging from 1 (minor doubts) to 3 (no
doubts).  These values were assigned based on the apparent shape of
the spectral profile and, if present, on the morphology and size of an
optical counterpart in the HST images.

\subsection{Lyman $\alpha$ emitter sample}
\label{sec:lyalpha-emitt-sample}

\begin{figure}
  \centering
  \includegraphics[width=0.46\textwidth]{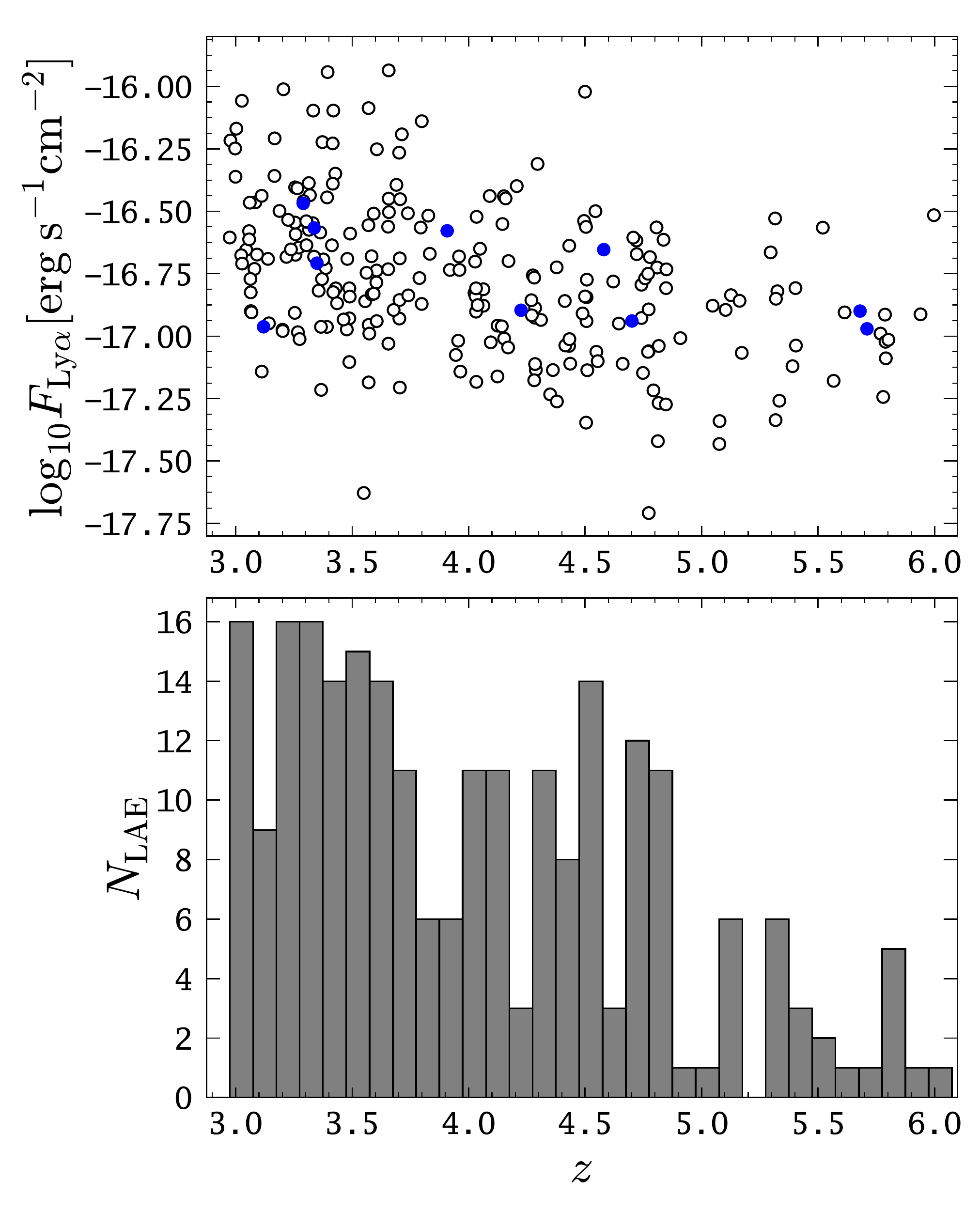}\vspace{-1em}
  \caption{\emph{Top panel}: Fluxes and redshifts of the MW LAE
    sample used in this study (\emph{open circles}) in
      comparison to the fluxes and redshifts of the MUSE HDFS LAEs
    used to determine a realistic selection function as described in
    Sect.~\ref{sec:select-funct-from-2} (\emph{filled
        circles}). \emph{Bottom panel}: Redshift histogram of
      the MW LAE sample (binning: $\Delta z = 0.1$).}
  \label{fig:flux_z_plot}
\end{figure}

The final emission line catalogue published in
\citetalias{Herenz2017a} consists of 831 emission line galaxies, with
237 Ly$\alpha$ emitting galaxies in the redshift range
$2.97 \leq z \leq 5.99$.  Two of these high-$z$ galaxies exhibit clear
signatures of active nuclei\footnote{MW IDs 104014050 and 115003085.} 
and are also flagged as active galaxies in the Chandra 7Ms source
catalogue \citep{Luo2017}.  Another object was classified as a strong
\ion{C}{iv} emitter, and is therefore also likely not a star-forming
LAE\footnote{MW ID 121033078.}.  We note that these AGN are also the
most luminous LAEs in our sample.  In our analysis below we 
discuss the effect of not excluding these {bona fide} AGNs when
determining the LAE LF.

All except five of the 234 non-AGN LAE galaxies have only a single line
detected by \texttt{LSDCat}.  The five exceptions are characterised by
strongly pronounced double-peaked Ly$\alpha$ profiles, with both peaks
having individual entries in the emission line catalogue\footnote{MW IDs
106014046, 115005089, 110003005, 122021111, and 123018120.}.  Moreover,
only 20 sources have confidence value 1 assigned, i.e. there were
minor doubts regarding their classification as Ly$\alpha$.  However, we
found that excluding those low-confidence sources from our analysis
had no impact on the resulting LF determinations.

Lyman $\alpha$ emitter redshifts  were determined by fitting the spectral profiles. As
detailed in \citetalias{Herenz2017a} we used the fitting formula
\begin{equation}
  \label{eq:asymeq}
  f(\lambda) = A \times \exp \left ( - 
     \frac{(\lambda - \lambda_0)^2}{2 \times (a_\mathrm{asym}(\lambda -
      \lambda_0) + d)^2} \right ) 
\end{equation}
introduced by \cite{Shibuya2014} to adequately model the asymmetric
spectral profiles of LAEs.  The free parameters $A$, $\lambda_0$,
$a_\mathrm{asym}$, and $d$ in Eq.~(\ref{eq:asymeq}) are the amplitude,
the peak wavelength, the asymmetry parameter, and the typical width of
the line, respectively.

Emission line fluxes $F_\mathrm{line}$ of the LAEs were determined
with the automated flux extraction routine of \texttt{LSDCat}.  In
\cite{Herenz2017} we found that for LAEs in the MW survey the
automatic measurements from the software compare best to a manual
curve-of-growth analysis over the spectral and spatial extent of the
emitters when aperture radii of three times the Kron-radius
\citep{Kron1980} were used.  Thus, we use these
$F_\mathrm{line}(3\times R_\mathrm{Kron})$ fluxes as the basis for our
luminosity function analysis.  The mean and median
$3\times R_\mathrm{Kron}$ radii in which fluxes were extracted are
2.1\arcsec{} and 2.0\arcsec{}, respectively, with values ranging from
1.8\arcsec{} to 3.7\arcsec{}.  However, we cautioned in
\citetalias{Herenz2017a} that quite frequently the spectral window of
the automated flux extraction did not completely encompass the whole
spectral profile of the LAEs.  These profiles are often characterised
by a weak secondary bump bluewards  of the main spectral peak.  This
may result in flux losses.  In order to correct for these losses, we
first visually inspected all spectral profiles to classify them as
single- or double-peaked.  We found that 90 LAEs in our sample show a
weaker secondary blue peak.  We then fitted all  double-peaked profiles
with a linear combination of two profiles described by
Eq.~(\ref{eq:asymeq}).  From these fits we calculated the fraction of
the line flux outside the spectral extraction window as flux
correction factor.  The average (median) flux correction factor for
the double-peaked emitters derived from this procedure is 1.17 (1.16).
Using the single component fits of Eq.~(\ref{eq:asymeq}) we also
derived flux correction factors for the single-peaked LAE profiles.
Here the correction factors are significantly smaller (mean: 1.03,
median: 1.02), thus indicating the overall robustness of the automated
procedure for simple emission line profiles.  The final LAE fluxes
used in our analysis are then obtained by multiplying the catalogued
$F_\mathrm{line}(3\times R_\mathrm{Kron})$ fluxes by each individually
determined correction factor\footnote{The catalogue of the LAE sample used in
    this publication is  available as an associated data product
    via the CDS.}.  An
overview of the fluxes and redshifts and a redshift histogram
of the MW LAE sample are shown in
Figure~\ref{fig:flux_z_plot}.

Finally, the measured fluxes are  converted into Ly$\alpha$
luminosities, 
\begin{equation}
  \label{eq:lyalum}
  L_\mathrm{Ly\alpha} = 4 \pi F_\mathrm{Ly\alpha} D_L^2(z)\;\text{,}
\end{equation}
where $D_L(z)$ is the luminosity distance corresponding to the
redshift of the Ly$\alpha$ emitter that was determined from fitting
the spectral line profile with Eq.~(\ref{eq:asymeq}).

\section{MUSE-Wide Lyman $\alpha$ emitter selection function}
\label{sec:constr-muse-wide}

\begin{figure*}[t!]
  \centering
  \includegraphics[width=\textwidth]{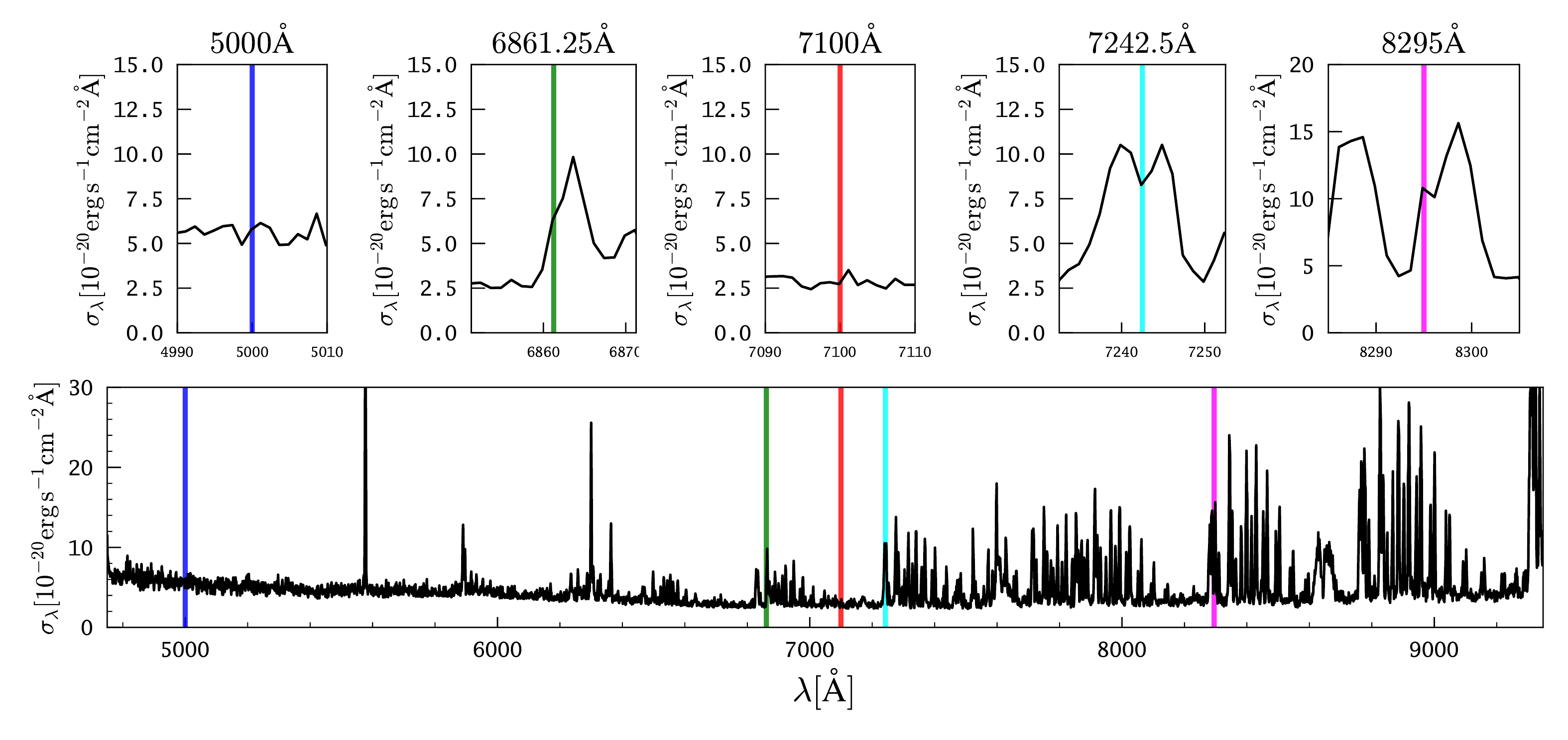}
  \vspace{-2em}
  \caption{Insertion wavelengths for completeness function estimation.
    The bottom panel shows the background noise over the whole
    spectral range; the vertical lines indicate the positions of the
    artificially implanted LAEs.  The top panels are zoomed-in
    versions around the regions of interest.  The colours of the
    vertical lines correspond to the colours used for the source
    recovery fractions in Figs.~\ref{fig:point_source01},
    \ref{fig:self}, and \ref{fig:selstack}.}
  \label{fig:efffsex}
\end{figure*}

\begin{figure}
  \centering
  \includegraphics[width=0.4\textwidth]{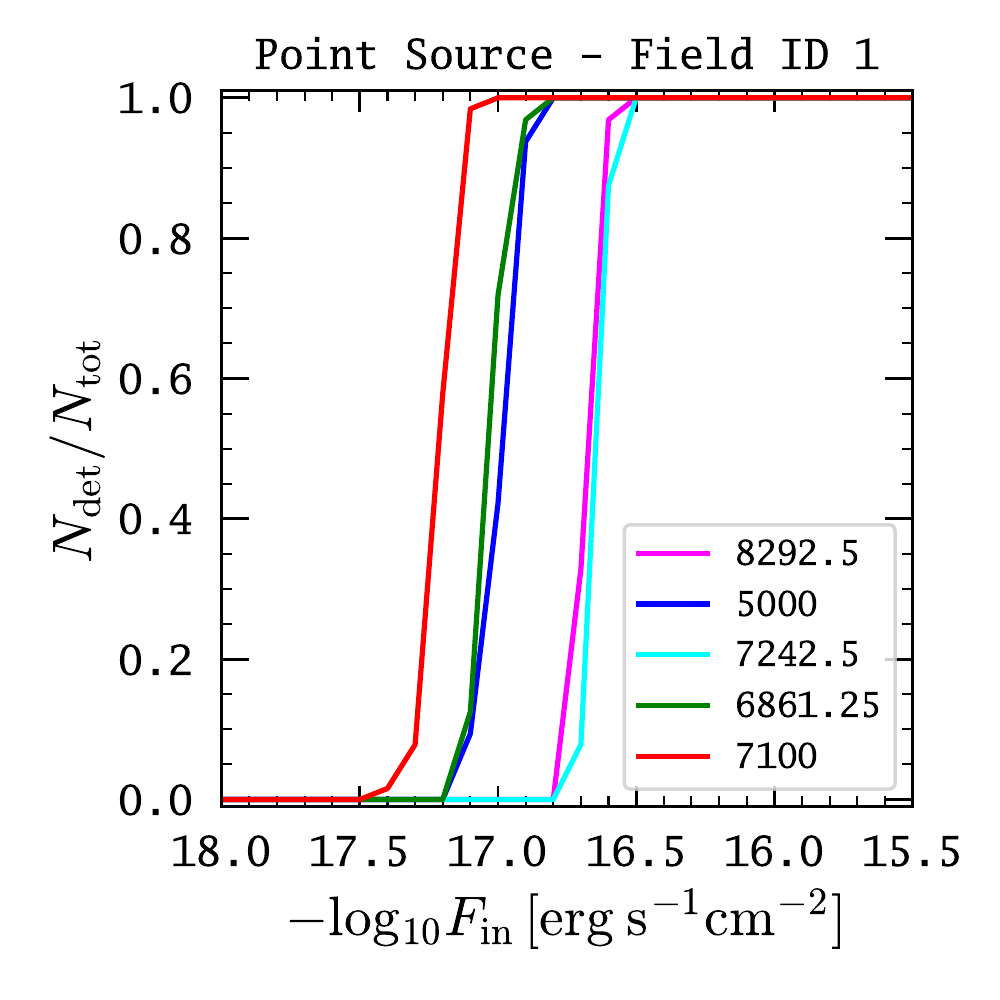}\vspace{-0.5em}
  \caption{Recovery fraction $N_\mathrm{det}/N_\mathrm{total}$ from a
    source insertion and recovery experiment for simplified point-like
    emission sources at five different wavelengths (see
    Figure~\ref{fig:efffsex}) in the MW pointing 01 datacube.
    $N_{total} = 64$ is the number of inserted sources at a given flux
    level and $N_\mathrm{det}$ is the number of recovered sources for
    a given line flux $F_\mathrm{in}$ obtained with same detection
    procedure used to construct the original MW catalogue.}
  \label{fig:point_source01}
\end{figure}

\begin{figure}
  \centering
  \includegraphics[width=0.5\textwidth]{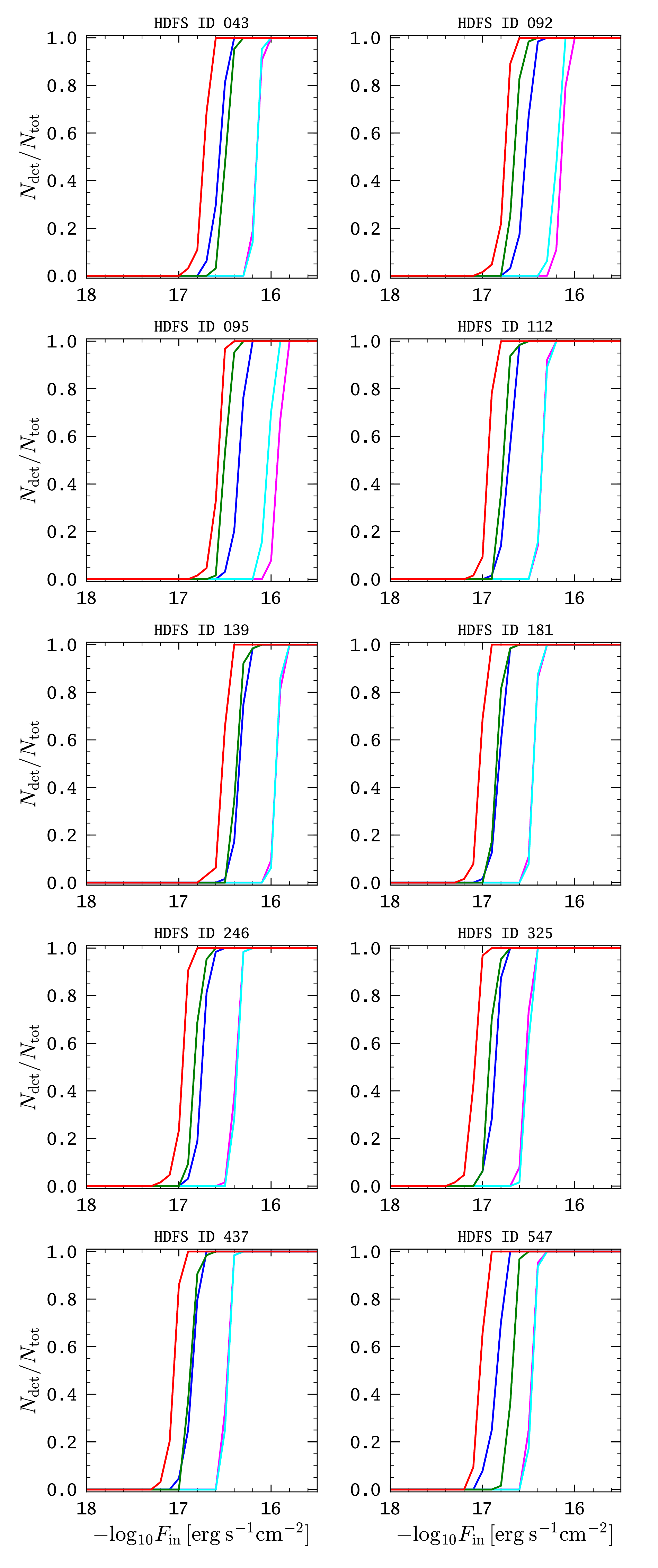}
  \vspace{-1.5em}
  \caption{Recovery fractions from a source insertion and recovery
    experiment with ten MUSE HDFS LAEs for MW datacube 01.  Each panel
    displays the recovery fraction $N_\mathrm{det}/N_\mathrm{total}$
    for a particular MUSE HDFS LAE  as a function of its scaled
    flux at five different wavelengths (see Figure~\ref{fig:efffsex}).
    $N_{total} $ and $N_\mathrm{det}$ are as defined in
    Figure~\ref{fig:point_source01}.  }
  \label{fig:self}
\end{figure}

\begin{figure}
  \centering
  \includegraphics[width=0.4\textwidth]{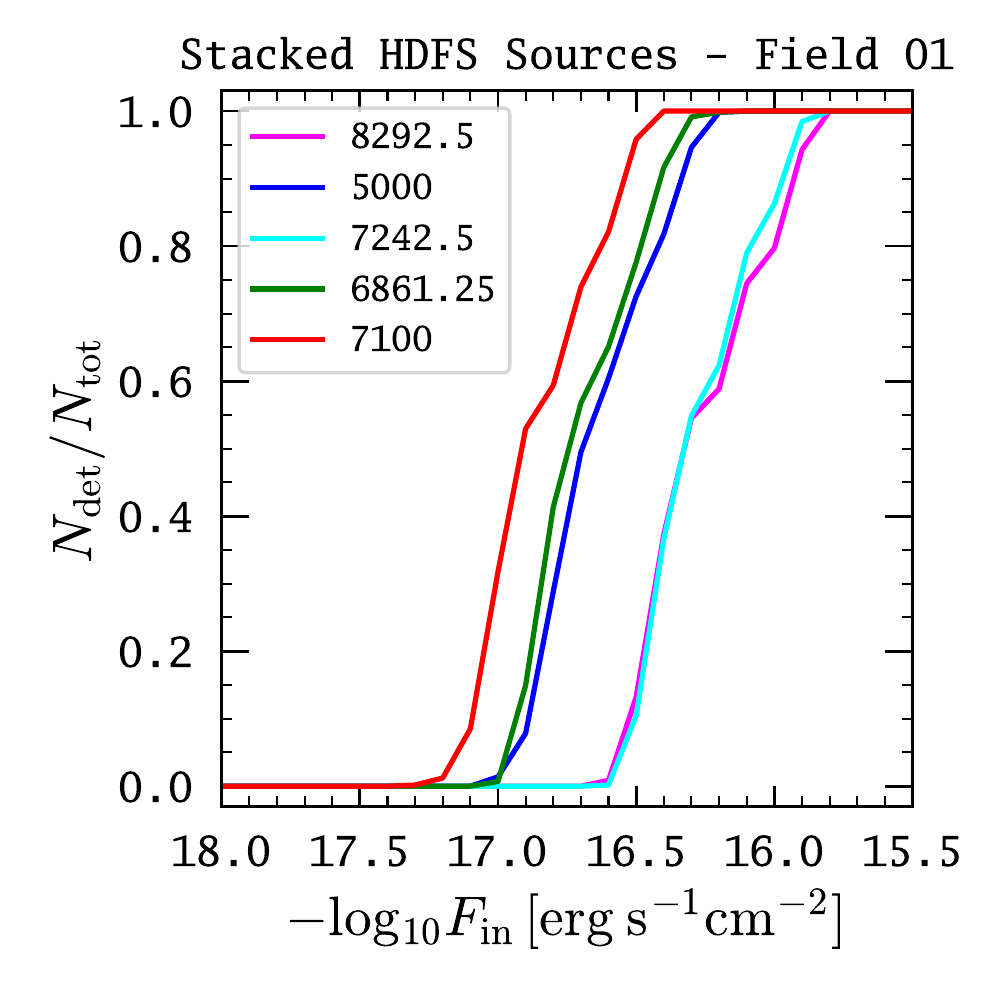}\vspace{-0.5em}
  \caption{Stack over the recovery fractions
    $N_\mathrm{det.}/N_\mathrm{total}$ of the ten different MUSE HDFS
    LAEs used in the source recovery experiment.  These curves
    represent the selection function at five different wavelengths in a
    MW datacube.  We only show  the results for the MW
    datacubes 01;  the shape of the curves are similar for
    all other fields.}
  \label{fig:selstack}
\end{figure}

To  derive the luminosity function from the MW LAE sample, we first
need to determine the selection function for LAEs in our
integral-field spectroscopic survey.  The selection function encodes
the probability $f_C(F_\mathrm{Ly\alpha}, \lambda)$ of observing a LAE
with flux $F_\mathrm{Ly\alpha}$ at wavelength $\lambda$ in our survey.
Given an adopted cosmology it can also be uniquely represented in
redshift-luminosity space:
$f_C(F_\mathrm{Ly\alpha}, \lambda) \leftrightarrow
f_C(L_\mathrm{Ly\alpha}, z_\mathrm{Ly\alpha})$.

In order to construct $f_C(F_\mathrm{Ly\alpha}, \lambda)$ for MW, we
studied the success rate of recovering artificially implanted LAEs with
our detection procedure.  In Sect.~\ref{sec:select-funct-from-1} we
discuss  this experiment as performed with model sources
characterised by a compact point-like spatial profile and a
  simple spectral profile.  Then, in
Sect.~\ref{sec:select-funct-from-2}, we discuss the experiment
as performed under more realistic assumptions by accounting for the
observed variety in spectral- and spatial profiles of LAEs.  To this
end we  make use of real LAEs observed in the MUSE HDFS
\citep{Bacon2015}.  Finally, we explain in
Sect.~\ref{sec:from-recov-fract} how the measured recovery fractions
are converted to selection functions
$f_C(F_\mathrm{Ly\alpha}, \lambda)$.

\subsection{Source recovery experiment with artificial point sources}
\label{sec:select-funct-from-1}

We first computed recovery fractions for an oversimplified case where
we assumed that LAEs are perfect point sources with simple spectral
profiles.  In particular we modelled the light profiles of the
implanted sources with a Moffat function \citep{Moffat1969}.  This
parameterisation provides a reasonably good approximation of the
seeing-induced PSF in ground-based optical to
near-infrared observations \citep{Trujillo2001}.  To account for the
wavelength dependence of the full width at half maximum (FWHM) of the
PSF, we used the coefficients of linear fits of
$\mathrm{FWHM}(\lambda)$ provided in Table 2 of
\citetalias{Herenz2017a}.  The spectral profile of the fake sources is
modelled as a simple Gaussian of 250 km\,s width (FWHM).

As it is computationally not feasible to perform the source insertion
and recovery experiment for all wavelength layers in each of the 24 MW
datacubes, we selected five insertion wavelengths that are
representative of typical noise situations in the datacube
(see Figure~\ref{fig:efffsex}): $\lambda_1=5000$\,\AA{},
$\lambda_2=6861.25$\,\AA{}, $\lambda_3=7100$\,\AA{},
$\lambda_4=7242.5$\,\AA{}, and $\lambda_5=8292.5$\,\AA{}.  In
particular, the spectral regions around 5000\AA{} and 7100\AA{} are
typical regions devoid of night sky line emission, while 6861.25\AA{}
is in the wing of a sky line, and the 7242.5\AA{} and 8292.5\AA{}
positions are chosen to be right between two neighbouring sky lines.
At these insertion wavelengths we then populate each of the 24 MW
cubes with $N_\mathrm{tot}=64$ fake sources at different spatial
positions.  Instead of placing the inserted sources on a regular grid,
we used a quasi-random grid based on a Sobol sequence \citep[see
e.g. Sect.~7.7 of][]{Press1992}.  This is done to avoid placement of
the sources at similar distances to the edges of the MUSE slice
stacks.  These stacks are arranged in a rectangular pattern, which is
only slightly modulated by the small dither offsets during the
observations.  With this procedure we ensured that our selection
function is not affected by systematic defects that are known to exist
at the slice stack edges \citep[see e.g. Fig.~3 in][]{Bacon2017}.  We
then inserted fake sources with 20 different flux levels from
$\log F_\mathrm{Ly\alpha} [\mathrm{erg\,s}^{-1}\mathrm{cm}^{-2}] =
-17.5$ to
$\log F_\mathrm{Ly\alpha} [\mathrm{erg\,s}^{-1}\mathrm{cm}^{-2}] =
-15.5$ in steps of 0.1\,dex at the five chosen wavelength layers into
each MW datacube.  The $20\times24=480$ datacubes were then continuum
subtracted with the running median filter as described in
Sect.~\ref{sec:lyman-alpha-emitter-1}.  We then process these
continuum subtracted cubes with \texttt{LSDCat} in the same way as for
the original catalogue construction
(Sect.~\ref{sec:lyman-alpha-emitter-1}).  In order to decrease the
computational cost for this experiment, we trimmed the continuum
subtracted fake-source populated datacubes by $\pm 30$\AA{} around
each insertion wavelength.  For each subcube we then counted the
number sources $N_\mathrm{det}$ that are recovered by \texttt{LSDCat}
above the same detection threshold ($\mathrm{S/N}_\mathrm{det} = 8$)
that was used for the creation of the MW emission line source
catalogue (see Sect.~\ref{sec:lyman-alpha-emitter-1}).  As an example,
we show in Fig.~\ref{fig:point_source01} the resulting recovery
fractions for each insertion wavelength for MW pointing 01.  We note
that the shape and order of the curves is similar for all other
pointings.

\subsection{Source recovery experiment with real LAEs}
\label{sec:select-funct-from-2}

We also performed a source insertion and recovery experiment using the
10 LAEs from the MUSE HDFS catalogue that have the highest S/N
(MUSE HDFS ID 43, 92, 95, 112, 139, 181, 246, 325, 437, and 547  all
have S/N$>$10). These sources show a range of different
surface-brightness profiles: e.g. while the LAEs 43, 92, and 95 are
fairly extended, the LAEs 181, 325, and 542 show more compact surface
brightness profiles \citep{Wisotzki2015}.  They also represent a range
of fluxes, redshifts, and line profiles.  Given their high S/N
in the MUSE HDFS data, they are practically noise free compared to the
noise level in MW, even when being multiplicatively rescaled to higher
flux levels.  We compare the fluxes and redshifts of these ten
LAEs to the actual MUSE-Wide sample in
  Figure~\ref{fig:flux_z_plot}.  It can be seen that all MUSE HDFS LAEs used
 in the source insertion experiment could potentially be part of the MW Sample.

We  rescaled these LAEs to 20 different flux levels between
$\log F_\mathrm{Ly\alpha} [\mathrm{erg\,s}^{-1}\mathrm{cm}^{-2}] =
-17.5$ to
$\log F_\mathrm{Ly\alpha} [\mathrm{erg\,s}^{-1}\mathrm{cm}^{-2}] =
-15.5$ in steps of 0.1\,dex (i.e. we used the same flux levels
  as  before for the simplified sources).  For this purpose we
first measured the fluxes from the MUSE HDFS LAEs by utilising
the \texttt{LSDCat} flux-measurement routine with circular apertures of
radius $3R_\mathrm{Kron}$.  We then cut out mini cubes from the MUSE
HDFS datacube that are centred on the LAEs.  The voxels in these
mini-cubes were then multiplied by constant factors to reach the
desired flux levels.  These 20$\times$10 (flux samples $\times$ source
samples) `fake-source' mini cubes were inserted into each of our 24
MW datacubes at the five different insertion wavelengths and at the
same positions that were  used for the simplified sources.

When inserting the sources at different wavelengths we accounted for
the redshift broadening of spectral profile, i.e. we kept the profile
shape fixed in velocity space.  We also needed to account for the
differences in the PSF between MW and MUSE HDFS.
Since in all MW datacubes the   PSF is broader
than it is in the HDFS, we have to degrade the PSF of the inserted
mini cubes. To this end we convolved their spatial layers with a 2D
Gaussian of dispersion
$\sigma_{2D}(\lambda) = \sqrt{\sigma_\mathrm{MW}(\lambda)^2 -
  \sigma_\mathrm{HDFS}(\lambda)^2}$, where
$\sigma_\mathrm{MW}(\lambda)$ and $\sigma_\mathrm{HDFS}(\lambda)$ are
the wavelength-dependent PSF dispersions of a MUSE-Wide datacube and
the MUSE HDFS datacube, respectively.  Here the MUSE HDFS PSF was
determined  from fits to the brightest star in the field (see
Fig. 2 of \citealt{Bacon2015}), while the linear model of
\citetalias{Herenz2017a} was used for the MW PSF.

After having continuum subtracted datacubes with artificially
implanted sources, the next step was to perform the recovery
experiment.  To reduce the computational cost of this experiment, we
trimmed the fake-source inserted cubes in wavelength range to
$\pm 30\AA{}$ around each insertion wavelength.  The full recovery
experiment was thus performed on
$20 \times 10 \times 5 \times 24 = 24000$ datacubes of dimensions
$\sim 300 \times 300 \times 50$ (neglecting empty edges due to the
rotation of the MW pointings).  Each of these cubes was processed with
\texttt{LSDCat} using the same parameters that were used to
generate the catalogue of LAEs in the 24 MW fields.  We then counted the
number of recovered sources $N_\mathrm{det}$ above the same detection
threshold that was used in the creation of the MW LAE source catalogue
($\mathrm{S/N}_\mathrm{det} = 8$).

We demonstrate the outcome of the recovery experiment with realistic
LAES for the MW pointing 01 datacube in Figure~\ref{fig:self}, noting
that the results for the other datacubes are similar. We found that
the completeness curves for all emitters have a very steep cut-off at
line fluxes of $10^{-16}$\dots$10^{-17}$\,erg\,s$^{-1}$cm$^{-2}$.
While for the more compact LAEs the cut-off is comparable to the one
obtained for the idealised sources
(cf. Figure~\ref{fig:point_source01}), for the more extended LAEs it
is significantly shifted to brighter flux levels.  The exact turnover
point on a given curve appears to be a complicated function of a
source's surface-brightness profile and its spectral profile.
However, we observe that for a given source all curves are
self-similar and the shift depends only on the insertion wavelength
(Fig.~\ref{fig:efffsex}).  Since the ten LAEs from the MUSE HDFS used
in the recovery experiment are expected to be a representative subset
of the overall high-$z$ LAE population, we expect the overall LAE
selection function at a specific wavelength to be the average recovery
fraction over all sources
$\langle N_\mathrm{det}/N_\mathrm{tot} \rangle $ at this wavelength.
In Fig.~\ref{fig:selstack} we show as an example these averaged
recovery fractions for MW pointing 01.  Similar to the idealised
sources, the shape and the order of the curves is similar for all
other pointings.

\subsection{From recovery fractions to selection functions}
\label{sec:from-recov-fract}

\begin{figure*}
  \centering
  \includegraphics[width=\textwidth]{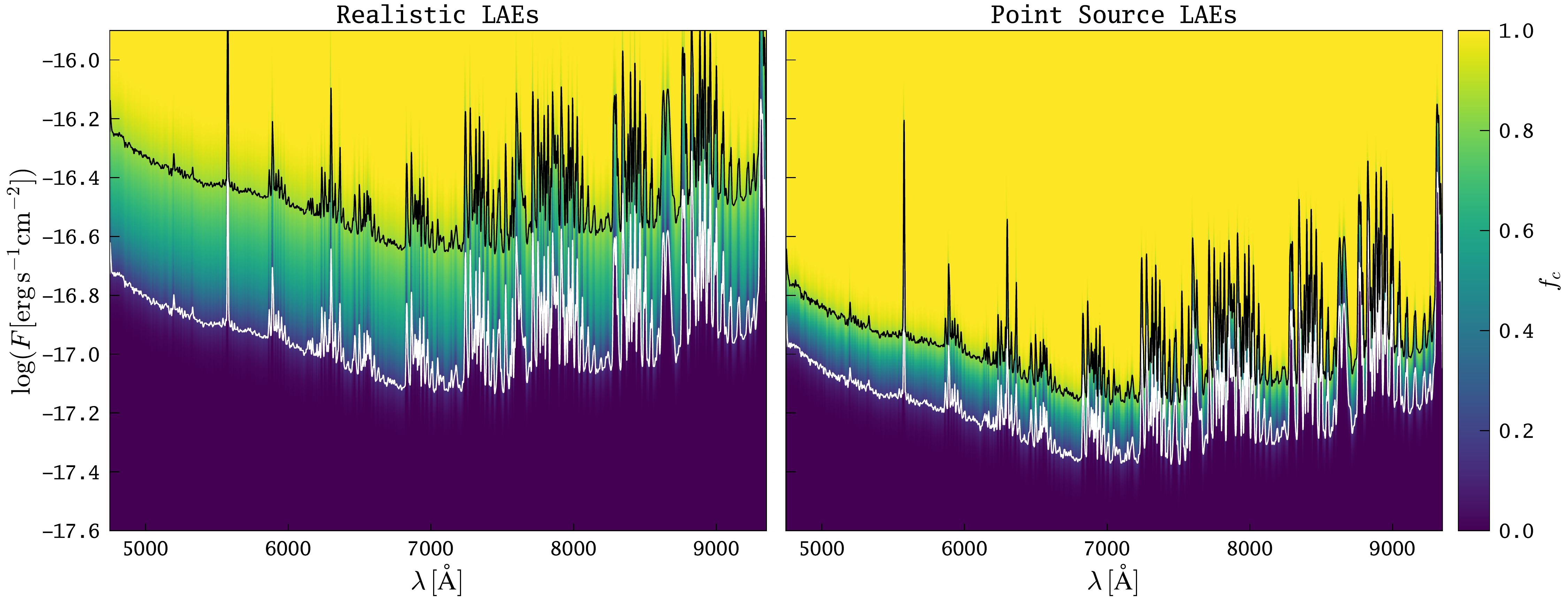}
  \caption{Selection function $f_C(F_\mathrm{Ly\alpha}, \lambda)$ for
    LAEs in the MW survey.  The white and black lines indicate
    the 15\% and 85\% completeness limits, respectively.  The left
    panel shows the RSSF (see
    Sect.~\ref{sec:select-funct-from-2}).  The right panel shows the
    PSSF (see
    Sect.~\ref{sec:select-funct-from-1}).}
  \label{fig:selfun}
\end{figure*}
\begin{figure*}
  \centering
  \includegraphics[width=\textwidth]{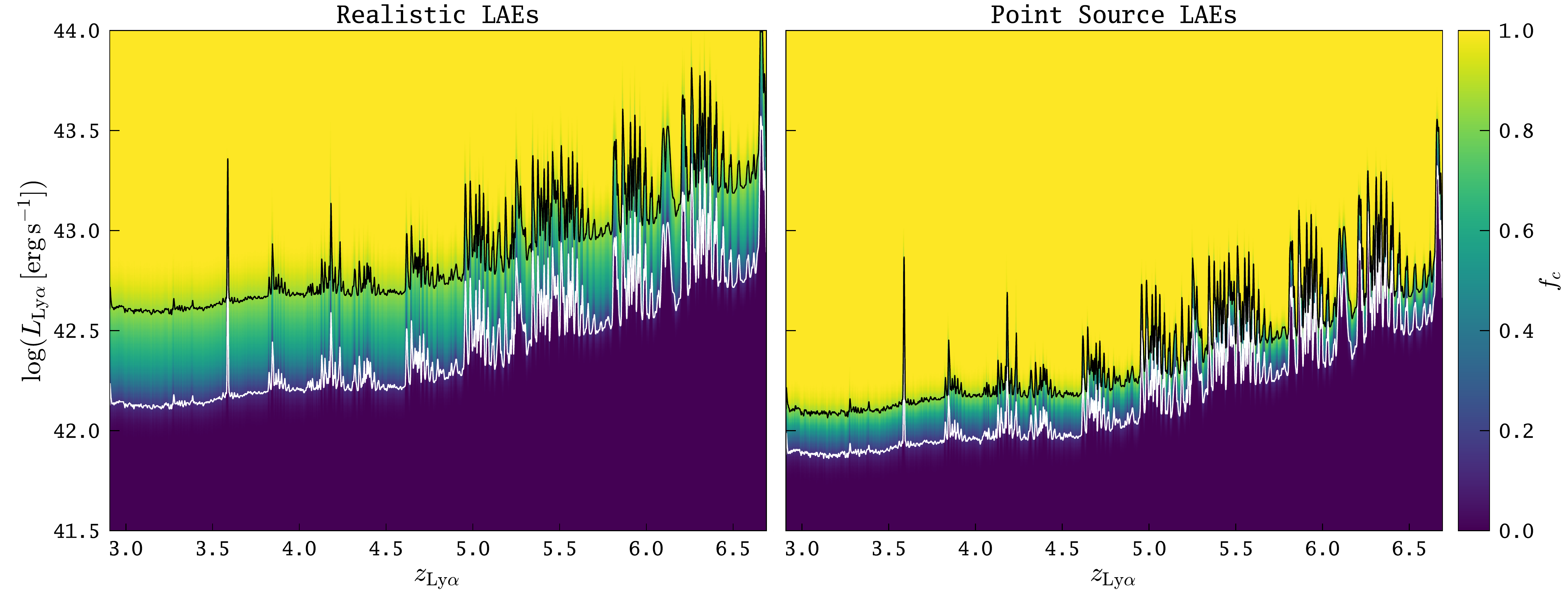}
  \caption{Selection function for LAEs in the MW survey,
    similar to Figure~\ref{fig:selfun}, but now transformed to
    redshift-luminosity space.}
  \label{fig:selfun_zlum}
\end{figure*}

Up to this point we are equipped with LAE selection functions for the
MW LAEs only at five different wavelengths within the MUSE wavelength
range.  However, we notice in Figure~\ref{fig:point_source01} and
Figure~\ref{fig:selstack} that the curves at the different wavelengths
are self-similar and that their order in flux is always the same.
This result indicates that there is a universally shaped selection
function whose shift with respect to the flux axis is determined by a
wavelength dependent quantity.  Indeed, we found that the shift of the
50\% completeness point ($f_{C}(F_{50}) = 0.5$) of the determined
curves shows a nearly constant $F_{50}$-to-$\tilde{\sigma}_\mathrm{emp}$
ratio for all curves, with $\tilde{\sigma}_\mathrm{emp}$ being the
empirically determined background noise convolved with a 250
km\,s$^{-1}$ wide (FWHM) Gaussian. The ratio
$F_{50}/\tilde{\sigma}_\mathrm{emp}(\lambda)$ varied between 400 and 460
for the different datacubes; the exact value depended on the average
datacube background noise and is a function of the observing
conditions.  Using this scaling we could compute
$f_{C,i}(F_\mathrm{Ly\alpha}, \lambda)$ for each of the 24 MW
pointings (here $i$ indexes the pointing). We created a master
$f(F)$-curve from shifting the five stacked curves on top of each other
by requiring them to have the same $f_{C}(F'_{50}) = 0.5$ value.  For
each wavelength bin we then shifted this $f(F)$-master curve according
to the $F_{50}/\tilde{\sigma}_\mathrm{emp}(\lambda)$ proportionality
to obtain $f_{C,i}(F_\mathrm{Ly\alpha}, \lambda)$.  The final
selection functions for the MW LAE catalogue were then the average of
all 24 selection functions.

The resulting selection function for the point-like emission line
sources is called the point source selection function (PSSF).  This
more realistic selection function is therefore called the real
source selection function (RSSF).  Both selection functions are
shown in Figure~\ref{fig:selfun} in redshift-flux
space\footnote{The LAE selection functions shown in
    Figure~\ref{fig:selfun} (RSSF and PSSF) are made available as 
    associated data products via the CDS.} and in
Figure~\ref{fig:selfun_zlum} redshift-luminosity space.

The PSSF can be seen as a limiting depth of our survey since it
resembles closely the template of the matched filter used in the
emission line source detection \citepalias{Herenz2017a}.  More
importantly, in comparison to the RSSF it also demonstrates the loss in
sensitivity in LAE surveys due to the fact that these sources are not
compact, but exhibit significant low surface brightness halo
components.  Moreover, while the transition from 0\% to 100\%
completeness is quite rapid for the PSSF, the variety of Ly$\alpha$
halo properties encountered amongst LAEs leads to a much smoother
transition.  Notably, in extreme cases Ly$\alpha$ halos can contain
up to 90\% of the total Ly$\alpha$ flux
\citep{Wisotzki2015,Leclercq2017}.  Therefore, the assumption of
point-like LAEs in estimating the selection function leads to an
overestimate of survey depth.  While \cite{Grove2009} already noted
this effect, they were not able to robustly quantify it due to the
lack of deeper comparison data.

\section{Deriving the Lyman $\alpha$ luminosity function}
\label{sec:lyman-alpha-emitter}

Before presenting the results of the LAE LF in the next section, we
provide here an overview of the methods used to derive the LAE LF in
our integral field spectroscopic dataset.

We use three different non-parametric LF estimators: the classical
$1/V_\mathrm{max}$ method (Sect.~\ref{sec:1v_mathrmmax-method}), a
binned alternative method to $1/V_\mathrm{max}$ introduced by
\citeauthor{Page2000} (Sect.~\ref{sec:phi_m-meth-citepp}), and the
$C^{-}$ method (Sect.~\ref{sec:c--method}).  As we  discuss, the
second and third method provide some advantages over the classical
$1/V_\mathrm{max}$ approach.  Moreover, we also make use of a
non-parametric method to test the redshift evolution of the LAE LF
(Sect.~\ref{sec:non-parametric-test}).
Furthermore, photometric uncertainties at low completeness levels will
lead to biases in the LF estimate. In order to avoid those biases we
truncate the sample and define appropriate luminosity bins for the
binned estimators.  We motivate our truncation criterion and bin-size
choice in Sect.~\ref{sec:binn-trunc-sample}.
Finally, in Sect.~\ref{sec:param-maxim-likel} we explain the
maximum-likelihood fitting formalism that we employ to derive
parametric models of the LAE LF.

\subsection{Non-parametric luminosity function estimates}
\label{sec:nn-param-lumin}

\subsubsection{The $1/V_\mathrm{max}$ method}
\label{sec:1v_mathrmmax-method}

The first non-parametric LF estimator we consider is the 
$1/V_\mathrm{max}$ estimator \citep{Schmidt1968,Felten1976} in a
modified version to account for a complex, i.e. redshift- and
luminosity-dependent, selection function \citep{Fan2001,Caditz2016}.

The $1/V_\mathrm{max}$ estimator approximates the cumulative
luminosity function
\begin{equation}
  \label{eq:2}
  \Phi(L_\mathrm{Ly\alpha}) = \int_\mathrm{L_{Ly\alpha}}^\infty
  \phi(L_\mathrm{Ly\alpha}')\,\mathrm{d}L_\mathrm{Ly\alpha}'\,\mathrm{,}
\end{equation}
where $\phi(L_\mathrm{Ly\alpha})$ is the differential LF introduced in
Eq.~(\ref{eq:1}), via
\begin{equation}
  \label{eq:3}
  \Phi(L_{\mathrm{Ly\alpha},k}) = \sum_{i \leq k}
  \frac{1}{V_{\mathrm{max},i}} \;\mathrm{.}
\end{equation}
Here, and in the following, we assume that the objects are ordered in
Ly$\alpha$ luminosity, i.e.
\begin{equation}
  \label{eq:22}
L_\mathrm{Ly\alpha,1} > L_\mathrm{Ly\alpha,2} > \dots >
L_\mathrm{Ly\alpha,N-1} > L_\mathrm{Ly\alpha,N}  \,\mathrm{.}
\end{equation}
In Eq.~(\ref{eq:3}) $V_{\mathrm{max},i}$  denotes the maximum volume
accessible for each LAE $i$ in the survey.  In the presence of our
redshift-dependent selection function $f_C(L, z)$
(Fig.~\ref{fig:selfun_zlum}) we can write
\begin{equation}
  \label{eq:4}
  V_{\mathrm{max},i} = \omega \int_{z_\mathrm{min}}^{z_\mathrm{max}}
  f_c(L_{\mathrm{Ly\alpha},i},z) \frac{\mathrm{d}V}{\mathrm{d}z}
  \,\mathrm{d}z 
\end{equation}
\citep[e.g.][]{Wisotzki1998,Johnston2011}.  Here
$\omega$ is the angular area of the survey
($\omega=22.2$\,arcmin$^2$ for the 24 fields of the first instalment
of the MW survey under consideration here),
$\frac{\mathrm{d}V}{\mathrm{d}z}$ is the differential cosmological
volume element\footnote{For a definition of
  $\frac{\mathrm{d}V}{\mathrm{d}z}$, see e.g.
  \cite{Hogg1999}.}, and $z_\mathrm{min}$ ($z_\mathrm{max}$)
denotes the lower (upper) limit of the redshift range under
consideration\footnote{In our study these limits are either imposed by
  the full spectral coverage of MUSE, i.e.
  $(z_\mathrm{min},z_\mathrm{max}) = (2.9,6.7)$, or by the redshift
  bins that we consider (see Table \ref{tab:taus} below).}.

Moreover, in the $1/V_\mathrm{max}$ formalism the differential LF can
be approximated by the binned estimator
\begin{equation}
  \label{eq:5}
  \phi_{1/V_\mathrm{max}}(\langle L_\mathrm{Ly\alpha} \rangle) = \frac{1}{\Delta
    L_\mathrm{Ly\alpha}} \sum_{k} \frac{1}{V_{\mathrm{max},k}} \;\mathrm{,}
\end{equation}
where $\langle L_\mathrm{Ly\alpha} \rangle$ is the average Ly$\alpha$
luminosity of a bin, $\Delta L_\mathrm{Ly\alpha}$ is the width of the
bin, and the sum runs over all sources $k$ in that bin.  The
uncertainty for each bin is defined as
\begin{equation}
  \label{eq:6}
  \Delta \phi_{1/V_\mathrm{max}} (\langle L_\mathrm{Ly\alpha} \rangle) =
  \sqrt{\frac{1}{\Delta L^2} \sum_i \frac{1}{V_{\mathrm{max},i}^2}}
\end{equation}
\cite[e.g.][]{Johnston2011}.

\subsubsection{Binned estimator proposed by \cite{Page2000}}
\label{sec:phi_m-meth-citepp}

The second non-parametric estimator we consider provides an
alternative binned estimate for the differential LF.  In its original
form it was proposed by \cite{Page2000}. Following \cite{Yuan2013},
who provide a thorough comparison with the $1/V_\mathrm{max}$ method,
we call it the $\phi_\mathrm{PC}$ estimator.  This estimator was
motivated by potential systematic biases in the $1/V_\mathrm{max}$
estimator close to the flux limit of the survey.  It has
not yet been utilised to derive LAE LFs.

Instead of considering the maximum volume accessible for each
individual source in the binned $1/V_\mathrm{max}$ estimator
(Eq.~\ref{eq:5}), \cite{Page2000} argue that it is more robust to
consider the average four-dimensional volume in redshift-luminosity
space for each bin and then to divide the number of sources present in
the bin by this hypervolume.  In the presence of a redshift dependent
selection function we can write the $\phi_\mathrm{PC}$ estimator as
\begin{equation}
  \label{eq:7}
  \phi_\mathrm{PC}(\langle L_\mathrm{Ly\alpha} \rangle) =
  \frac{N}{\omega \int_{L_\mathrm{min}}^{L_\mathrm{max}}
    \int_{z_\mathrm{min}}^{z_\mathrm{max}}
    f_c(L_{\mathrm{Ly\alpha}},z) \, \frac{\mathrm{d}V}{\mathrm{d}z} \,
    \mathrm{d}z \, \mathrm{d}L} \; \mathrm{,}
\end{equation}
where again $\langle L_\mathrm{Ly\alpha} \rangle$ denotes the average
Ly$\alpha$ luminosity of a bin, $z_\mathrm{min}$ and $z_\mathrm{max}$
are the limits of the redshift range under consideration,
$L_\mathrm{min}$ and $L_\mathrm{max}$ are the lower and upper bounds of
the bin in which the LF is estimated, and $N$ is the number of sources
within the bin.  In analogy to Eq.~(\ref{eq:6}), we estimate the
statistical uncertainty on 
$\phi_\mathrm{PC}(\langle L_\mathrm{Ly\alpha} \rangle)$ by replacing
$N$ with $\sqrt{N}$ in Eq.~(\ref{eq:7}).

\subsubsection{The $C^{-}$ method}
\label{sec:c--method}

We also consider the $C^{-}$ method for estimating the cumulative LF
defined in Eq.~\ref{eq:2}.  This method was introduced into the
astronomical literature by \cite{Lynden-Bell1971} and the
generalisation for complex selection functions was introduced by
\cite{Petrosian1992}. To date, the generalised $C^{-}$ method has
not been used to derive LAE LFs.  Formal derivations of the method in
the presence of a redshift- and luminosity-dependent selection
function are given elsewhere
\citep[e.g.][]{Fan2001,Johnston2011,Caditz2016}; here we just summarise
the computational algorithm\footnote{An introduction into the $C^{-}$
  method is also presented in Chapter~4.9.1. of the \cite{Ivezic2014}
  textbook.}.

The first step in the generalised $C^{-}$ method is to define the
generalised comparable set $J_i$ for each LAE $i$ that contains all
LAEs $j$ with higher Ly$\alpha$ luminosity:
\begin{equation}
  \label{eq:8}
  J_i = \{j: L_j > L_i \} \;\mathrm{.}
\end{equation}
The next step is to make a weighted count of the number of LAEs in
each comparable set
\begin{equation}
  \label{eq:9}
  T_i = \sum_{j=1}^{N_i} w_j \;\mathrm{,}
\end{equation}
where $N_i$ is the number of LAEs in the comparable set $J_i$.  The
weights $w_j$ for each object $j$ in $J_i$ are given by the selection
probability if the $J_i$-defining object $i$ with its Ly$\alpha$
luminosity $L_{\mathrm{Ly\alpha},i}$ had  been detected at the
redshift of an object $j$, $f_c(L_{\mathrm{Ly\alpha},i},z_j)$,
normalised by $j$'s actual selection probability
$f_c(L_{\mathrm{Ly\alpha},i},z_j)$, i.e.
\begin{equation}
  \label{eq:12}
  w_j =
  \frac{f_c(L_{\mathrm{Ly\alpha},i},z_j)}{f_c(L_{\mathrm{Ly\alpha},j},z_j)}
    \; \mathrm{.}
\end{equation}
Since by construction
$L_{\mathrm{Ly\alpha},j} > L_{\mathrm{Ly\alpha},i}$, and since $f_c$
is monotonically increasing with luminosity at a given redshift,
$w_j \leq 1$ holds.  With these weighted counts  the cumulative
LAE LF is given as
\begin{equation}
  \label{eq:10}
  \Phi(L_\mathrm{Ly\alpha,k}) = \Phi(L_\mathrm{Ly\alpha,1})
  \prod_{i=2}^{k} \left ( 1 + \frac{1}{T_i} \right) \; \mathrm{,}
\end{equation}
where the normalisation $\Phi(L_\mathrm{Ly\alpha,1})$ has to be
determined separately (see Sect.~\ref{sec:non-parametric-test} below).

A potential advantage of the $C^{-}$ method over the
$1/V_\mathrm{max}$ method is that it only requires evaluation of the
selection function at redshifts where  sources were actually detected,
whereas the calculation of the LF using the $1/V_\mathrm{max}$ method
 requires integration over the selection function for the whole
 redshift range of interest.
 
\cite{Caditz2016} provides a detailed formal comparison between the
$C^{-}$ and $1/V_\mathrm{max}$ estimators, showing that both are
asymptotically unbiased, i.e. both $1/V_\mathrm{max}$ and $C^{-}$
yield a correct estimate of the true luminosity function for large
number of objects and a correct estimate of the selection function.
However, the main difference between the two estimators is that
$1/V_\mathrm{max}$ is more sensitive to uncertainties in the selection
function, while $C^{-}$ is more sensitive to random fluctuations in
the sample.

\subsubsection{Non-parametric test for LF evolution}
\label{sec:non-parametric-test}

\begin{figure*}
  \centering
  \includegraphics[width=\textwidth]{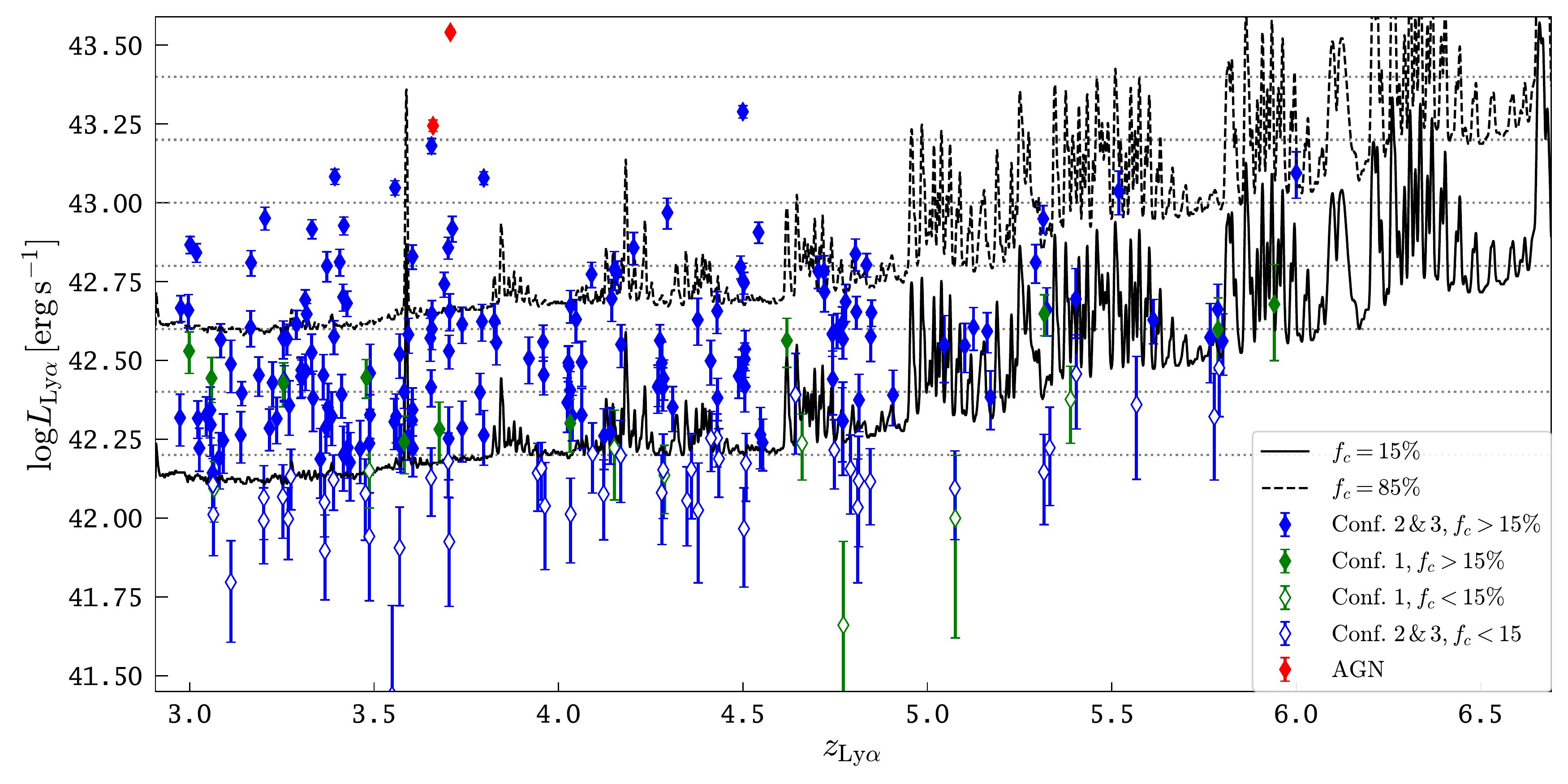}
  \vspace{-1.7em}
  \caption{LAE sample of the first 24 MW pointings in
    redshift-luminosity space.  The dashed line represents the 85\%
    RSSF completeness limit, while the black line denotes the 15\%
    RSSF completeness limit at which we truncate our sample leaving  179 of
    237 (75.6\%) LAEs.  Individual
    emitters are colour-coded according to their assigned confidence
    flags (blue:  little to no doubt on being an LAE; green: 
    LAEs flagged as uncertain;  for details on assigning the confidence
    values, see Sect.~3.2 of
    \citetalias{Herenz2017a}). The two highest $L_\mathrm{Ly\alpha}$
    objects are AGN indicated by red symbols.  Sources below the
    truncation line are shown with open symbols.  Horizontal dotted
    lines denote the adopted bin boundaries
    ($\log L_\mathrm{Ly\alpha,bin} [\mathrm{erg\,s^{-1}}] = 42.2 +
    i\times 0.2$ for $i=0,1,\dots,5$) for the binned LAE LF
    estimates.}  
  \label{fig:sources_selfun_zlum}
\end{figure*}

The $1/V_\mathrm{max}$ method as formulated in
Sect.~\ref{sec:1v_mathrmmax-method} explicitly assumes that the LF is
non-evolving over the redshift range under consideration, whereas the
key assumption in the  $C^{-}$ method described above is that the
distribution function $\Psi(L,z)$, which describes a potentially evolving
LF as a scalar field in redshift-luminosity space, is separable, i.e.
\begin{equation}
  \label{eq:11}
  \psi(L,z) = \rho(z) \phi(L) \;\mathrm{.}
\end{equation}
Here $\rho(z)$ describes the  mean density of sources as a function of
redshift.  Thus, if Eq.~(\ref{eq:11}) is an adequate description of the
evolving LF, then $\phi(L)$, and correspondingly $\Phi(L)$, would
retain its shape over the redshift range under consideration, with
only the overall normalisation being allowed to change.

The assumption of an LF evolving according to Eq.~(\ref{eq:11}) is
commonly referred to as pure density evolution.  In principle, $\rho(z)$
can also be determined with the formalism described above, by just
exchanging redshifts with luminosities of object $i$ in
Eq.~(\ref{eq:9}) and then using Eq.~(\ref{eq:10}) to estimate
$\rho(z)$.  While such a derivation could be used to normalise the
cumulative LF from the $C^{-}$ method,  here we take the shortcut by
utilising $\Phi(L_\mathrm{Ly\alpha,1})$ from the $1/V_\mathrm{max}$
method in Eq.~(\ref{eq:10}),
\begin{equation}
  \label{eq:15}
  \Phi(L_\mathrm{Ly\alpha,1}) = \frac{1}{V_{\mathrm{max},1}} \;\text{,}
\end{equation}
i.e. we implicitly assume that
$\rho$ is constant over the redshift ranges under consideration.

Following \cite{Fan2001}, we test the validity of the pure density
evolution of the LAE LF in the luminosity range probed by our survey
with the statistical test developed by \cite{Efron1992}.  We
calculate for each $J_i$ the generalised rank $R_i$ of $z_i$:
\begin{equation}
  \label{eq:13}
  R_i = \sum_{j=1}^{N_i} w_j \Theta (z_i - z_j) \;\text{, with} \;
  \Theta (x) =
  \begin{cases}
    0\;\text{for}\;x<0 \\
    1\;\text{for}\;x\geq 0 
  \end{cases} \mathrm{.}
\end{equation}
If $z$ is independent of $L$ in the sense of Eq.~(\ref{eq:11}), then
the $R_i$ values should be distributed uniformly between $0$ and the
corresponding $T_i$ values, i.e. the expectation value of $R_i$ is
$E_i = T_i/2$ and its variance is $V_i = T_i^2/12$.  Moreover, 
the statistic
\begin{equation}
  \label{eq:14}
  \tau = \frac{\sum_{i} (R_i - E_i)}{\sqrt{\sum_i V_i}}
\end{equation}
is approximately a standard normal distribution under the null
hypothesis that independence between $z$ and $L$ in Eq.~(\ref{eq:11})
is valid.

We follow the literature by adopting $|\tau| < 1$ as the critical
value at which the independence assumption cannot be rejected
\citep[][]{Efron1992,Fan2001}.  We point out that for a standard normal
distribution this value corresponds to p-values $p_0>0.16$, i.e. it is
decidedly higher than commonly adopted significance levels to reject
the null hypothesis (e.g. $p_0<0.05$ for $1\sigma$).

\subsection{Truncation and binning of the sample}
\label{sec:binn-trunc-sample}

Non-parametric estimates of the differential luminosity function,
regardless of the  estimator used, require binning of the sample in
luminosities.  Moreover, at the faintest luminosities the photometric
uncertainties become so large that they would translate into a large
uncertainties for the completeness correction in the LF estimation.
This potential bias can be avoided by trimming the sample from such
sources.  We visualise our choice of bin sizes and truncation limit
for the RSSF in Figure~\ref{fig:sources_selfun_zlum}.

We curtail the sample from sources that are detected below the
$f_c = 0.15$ completeness limit.  As can be seen in
Figure~\ref{fig:sources_selfun_zlum}, the vast majority of LAEs below
the $f_c = 0.15$ limit have photometric errors that extend below the
0\% completeness line, which provides the main motivation for this
truncation limit.  This truncation limit removes 54 LAEs from the
initial MW LAE sample for the RSSF.  In the calculation of the
luminosity function, we account for the truncation limit by setting
$f_c \equiv 0$ for all $f_c < 0.15$.

We  chose our lowest luminosity boundary to be
$\log L_\mathrm{Ly\alpha} [\mathrm{erg\,s}^{-1}] = 42.2$, motivated
 by the fact that it straddles our RSSF truncation
criterion in the $z\lesssim 5$ region in the sample
(Figure~\ref{fig:sources_selfun_zlum}).  However, as we opt for an
integer single digit, this removes four additional objects from the LF
sample truncated according to the RSSF.  For the PSSF all sources except one
have $f_c > 0.15$ above
$\log L_\mathrm{Ly\alpha} [\mathrm{erg\,s}^{-1}] = 42.2$.  We chose
our adopted bin-size
$\Delta \log L_\mathrm{Ly\alpha} [\mathrm{erg\,s}^{-1}] = 0.2$ because it is significantly larger than the photometric error in the lowest
luminosity bin.  Moreover, we  show in
Sect.~\ref{sec:parametric-results} that for this bin-size the
non-parametric estimates are in optimal agreement with the parametric
maximum-likelihood solution.

Although estimating the binned differential LF is popular in the
literature, we point out that binning represents a loss of
information\footnote{A recent discussion of the pitfalls when
  using binning in the analysis of astronomical data was presented
  in \cite{Steinhardt2018}.}, while all information present in the
source catalogue is retained when deriving the cumulative LF
\citep{Felten1976,Caditz2016}.  Moreover, the adopted
maximum-likelihood procedure (see Sect. 4.3) does not
require binning of the data.   Here we use the binned estimates only
for visual comparison to the binned values from the literature in
combination with our derived Schechter parameterisation
(see Sect.~\ref{sec:discussion}).

\subsection{Parametric maximum likelihood luminosity function
  estimation}
\label{sec:param-maxim-likel}

In order to obtain a parametric description of the MW LAE LF we use
the maximum likelihood parameter estimation approach introduced by
\cite{Sandage1979} into the field of observational cosmology.  Maximum
likelihood estimation is a statistical technique used to estimate the
parameters of a model given the data. We therefore need to assume an
analytical expression for the LF.  The Schechter function
\citep{Schechter1976} is the most commonly adopted functional form for
the Ly$\alpha$ LF:
\begin{equation}
  \label{eq:schechter}
  \phi(L)\,\mathrm{d}L = \phi^* \left ( \frac{L}{L^*} \right
  )^{\alpha} \exp \left ( - \frac{L}{L^*} \right ) \,
    \frac{\mathrm{d}L}{L^*} \; \text{.}
\end{equation}
We obtain the free parameters $L^*$ (characteristic luminosity in
erg\,s$^{-1}$), $\alpha$ (faint-end slope), and $\phi^*$ (normalisation
in Mpc$^{-3}$) by maximising the likelihood function
\begin{equation}
  \label{eq:18}
  \mathcal{L} = \prod_{i=1}^{N_\mathrm{LAE}} p(L_i,z_i)\;\mathrm{,}
\end{equation}
where
\begin{equation}
  \label{eq:19}
  p(L_i,z_i) = \frac{\phi(L_i) f_c(L_i,z_i)}{\int_{L_\mathrm{min}}^{L_\mathrm{max}} \int_{z_\mathrm{min}}^{z_\mathrm{max}}
    \phi(L) f_c(L,z) \frac{\mathrm{d} V}{\mathrm{d} z}\, \mathrm{d}L\,\mathrm{d}z}
\end{equation}
\citep[e.g.][]{Sandage1979,Fan2001,Johnston2011}.  In practice we search for the minimum of
\begin{equation}
  \label{eq:17}
  S = -2 \times \ln \mathcal{L} \;\text{.}
\end{equation}
Evaluation of this equation thus requires a summation over the entire
unbinned sample.  As can be seen in Eq.~(\ref{eq:19}), the space density
scaling factor $\phi^*$ cancels out and is thus not really a free
parameter in the fitting process. For any given combination of $L^*$
and $\alpha$ the value of $\phi^*$ is, however, uniquely determined
since the integral in the denominator must equal the total number of
objects in the sample used to calculate the likelihood function
\citep[e.g.][]{Mehta2015}.

Even simpler than a Schechter function is a power-law distribution of
\begin{equation}
  \label{eq:16}
\phi(L)\,\mathrm{d}L  = \frac{\phi^*}{L^*} \times L^\beta \, \mathrm{d}L\;  \text{,}
\end{equation}
which lacks the exponential cut-off and thus implies a larger fraction
of high-luminosity objects for equal power-law indices
$\beta = \alpha$.  Comparing Eq.~(\ref{eq:16}) to Eq.~(\ref{eq:19}) it
becomes evident that only $\beta$ is a free parameter in the
likelihood function, but as above the ratio $\phi^*/L^*$
is uniquely constrained by the total number of objects.

We do not consider more complex parametric expressions for the
Ly$\alpha$ LF such as a double power law because, as demonstrated
in Sect.~\ref{sec:parametric-results}, they are not required for
our data.

\section{Results of the Lyman $\alpha$ luminosity function}
\label{sec:results}

\subsection{Non-parametric reconstructions of the LAE LF}
\label{sec:non-param-results}

\begin{table}
  \caption{Results of statistical test according to Eq.~(\ref{eq:14})
    for testing the assumption of pure density evolution as defined in
    Eq.~(\ref{eq:11}).  }
  \label{tab:taus}
  \centering
  \begin{tabular}{lcccc} \hline \hline \noalign{\smallskip} 
    Redshift range & $|\tau_\mathrm{PSSF}|$ & $|\tau_\mathrm{RSSF}|$ & $p_{\mathrm{PSSF}}$ & $p_{\mathrm{RSSF}}$ \\
    \hline \noalign{\smallskip}
    $2.9 < z \leq 4$   & 0.47 & 0.24   & 0.32  &  0.40   \\ \noalign{\smallskip}
    $4.0 < z \leq 5.0$ & 0.79 & 0.98   & 0.21  &  0.16   \\ \noalign{\smallskip}
    $5.0 < z \leq 6.7$ & 0.05 & 0.29   & 0.48  &  0.39   \\ \noalign{\smallskip} \hline \noalign{\smallskip} 
    $2.9 < z \leq 6.7$ & 0.46 & 0.31   & 0.32  &  0.38   \\ \hline \hline \noalign{\smallskip}
  \end{tabular} \vspace{1em}
  \tablefoot{$\tau$ values were computed for the point
    source selection function ($\tau_\mathrm{PSSF}$) and the selection
    function accounting for extended Ly$\alpha$ emission
    ($\tau_\mathrm{RSSF}$).  The corresponding  values of p for a standard
    normal distribution are given in Cols. 3 and 4.}
\end{table}

\begin{table*}
  \caption{Binned differential LAE LF from the first 24 MW pointings.}
  \label{tab:bin_lf}
  \centering
  \begin{tabular}{cccccc}
    \hline \hline  \noalign{\smallskip}
    $\log L_\mathrm{Ly\alpha}$ & $N_\mathrm{LAE}$ & $\phi_\mathrm{PC}$  &  $\Delta \phi_\mathrm{PC}$  & $\phi_\mathrm{1/V_\mathrm{max}}$  & $\Delta \phi_\mathrm{1/V_\mathrm{max}}$ \\
    (erg\,s$^{-1}$) & {} & (Mpc$^{-3} [\Delta \log L_\mathrm{Ly\alpha}]^{-1}$) &(Mpc$^{-3} [\Delta \log L_\mathrm{Ly\alpha}]^{-1}$) & (Mpc$^{-3} [\Delta \log L_\mathrm{Ly\alpha}]^{-1}$)& (Mpc$^{-3} [\Delta \log L_\mathrm{Ly\alpha}]^{-1}$) \\ \noalign{\smallskip}\hline \noalign{\smallskip} 
    42.3 & 52& $5.5 \times 10^{-3}$  & $7.7 \times 10^{-4}$ & $5.9 \times 10^{-3}$  & $8.6 \times 10^{-4}$ \\
    42.5 & 59& $3.0 \times 10^{-3}$  & $4.0 \times 10^{-4}$ & $3.1 \times 10^{-3}$  & $4.1 \times 10^{-4}$ \\
    42.7 & 40& $1.4 \times 10^{-3}$  & $2.2 \times 10^{-4}$ & $1.4 \times 10^{-3}$  & $2.3 \times 10^{-4}$ \\
    42.9 & 17& $4.7 \times 10^{-4}$  & $1.1 \times 10^{-4}$ & $4.8 \times 10^{-4}$  & $1.2 \times 10^{-4}$ \\
    43.1 & 6& $1.4 \times 10^{-4}$  & $5.8 \times 10^{-5}$ & $1.5 \times 10^{-4}$  & $5.9 \times 10^{-5}$ \\
    43.3 & 1& $2.3 \times 10^{-5}$  & $2.3 \times 10^{-5}$ & $2.3 \times 10^{-5}$  & $2.3 \times 10^{-5}$ \\ \noalign{\smallskip}\hline \hline \noalign{\smallskip} 
  \end{tabular}
  \tablefoot{$\phi_\mathrm{PC}$ is computed with the \cite{Page2000}
    estimator (Sect.~\ref{sec:phi_m-meth-citepp}), while
    $\phi_{1/V_\mathrm{max}}$ results from the binned $1/V_\mathrm{max}$
    estimator (Sect.~\ref{sec:1v_mathrmmax-method}).}
\end{table*}

\begin{figure}
  \centering
  \includegraphics[width=0.5\textwidth]{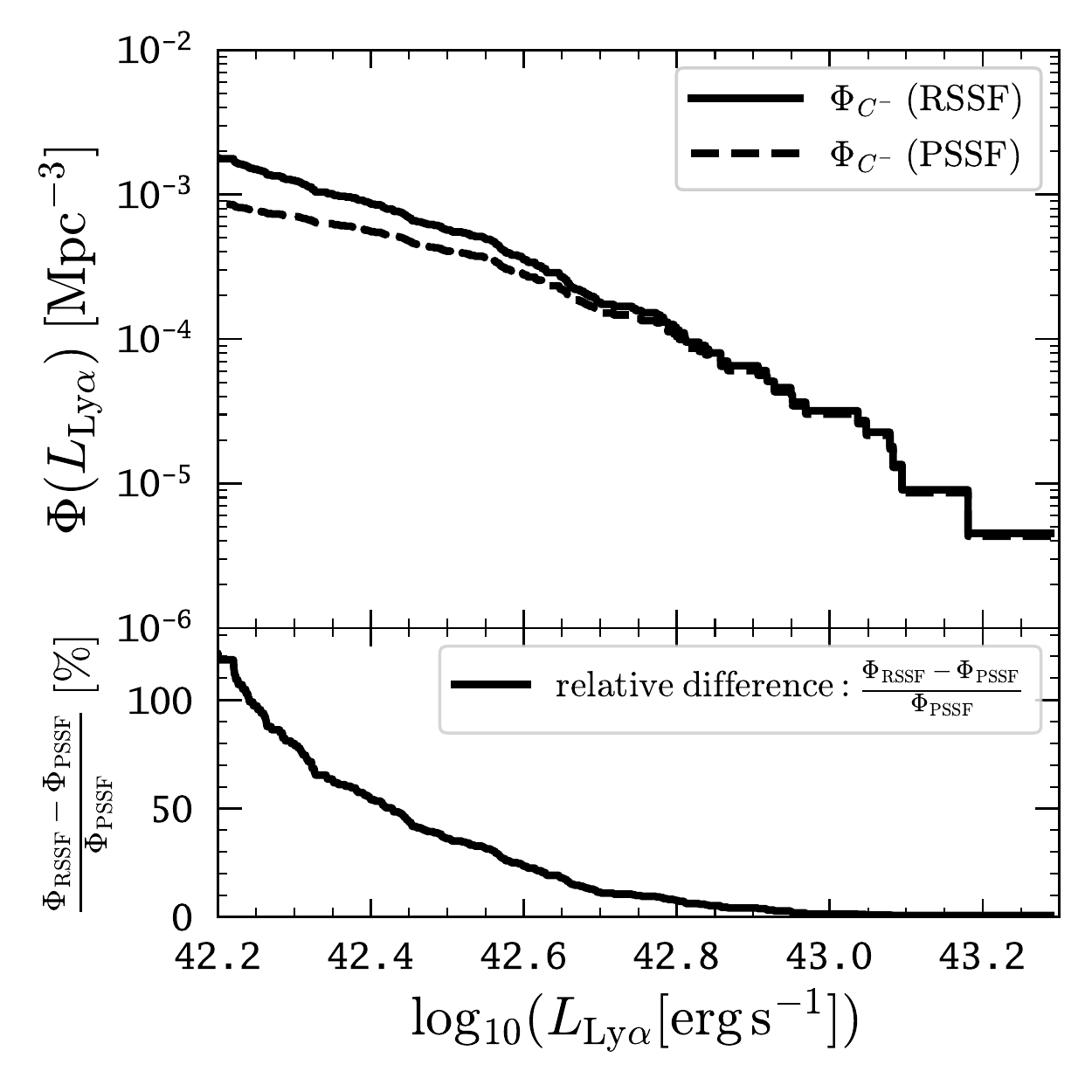}\vspace{-1em}
  \caption{\emph{Top panel}: Global ($2.9 \leq z \leq 6.7$) cumulative
    LAE LF from MW obtained with the $C^{-}$ method utilising
    the RSSF and the PSSF. \emph{Bottom panel}: Relative
    difference (\%)  between cumulative LFs utilising the RSSF
    and PSSF.}
  \label{fig:rssfpssfcomp}
\end{figure}

\begin{figure}
  \centering
  \includegraphics[width=0.5\textwidth]{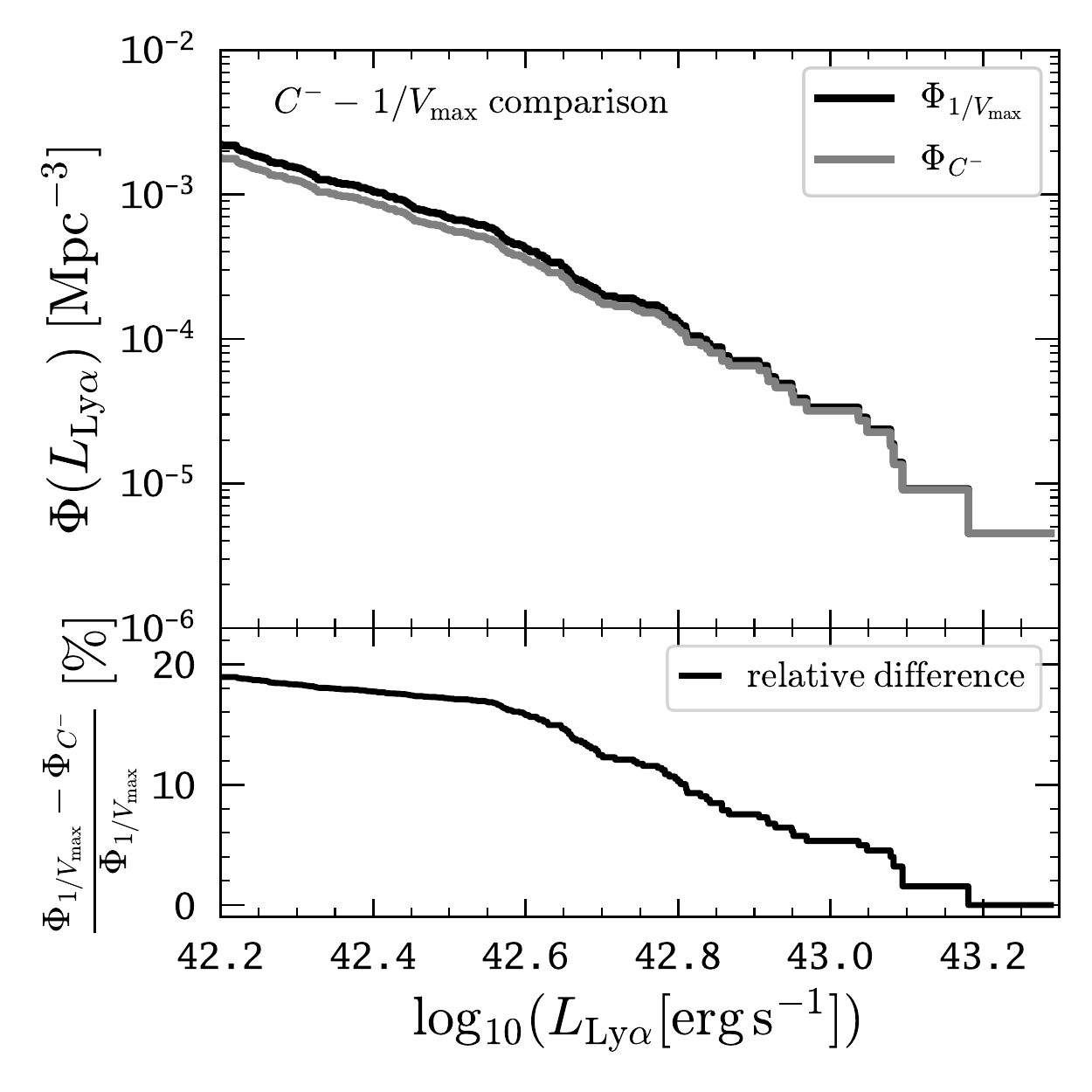}\vspace{-1em}
  \caption{ Comparison of the RSSF completeness corrected cumulative
    LAE LFs obtained with the $\mathrm{C}^{-}$ and
    $1/V_\mathrm{max}$ estimators.  \emph{Top panel}:  Cumulative LAE LFs
    from the two methods. \emph{Bottom panel}: Relative difference (\%)\ between the $1/V_\mathrm{max}$ and $\mathrm{C}^{-}$ methods.}
  \label{fig:globla_comp}
\end{figure}

\begin{figure}
  \centering
  \includegraphics[width=0.5\textwidth]{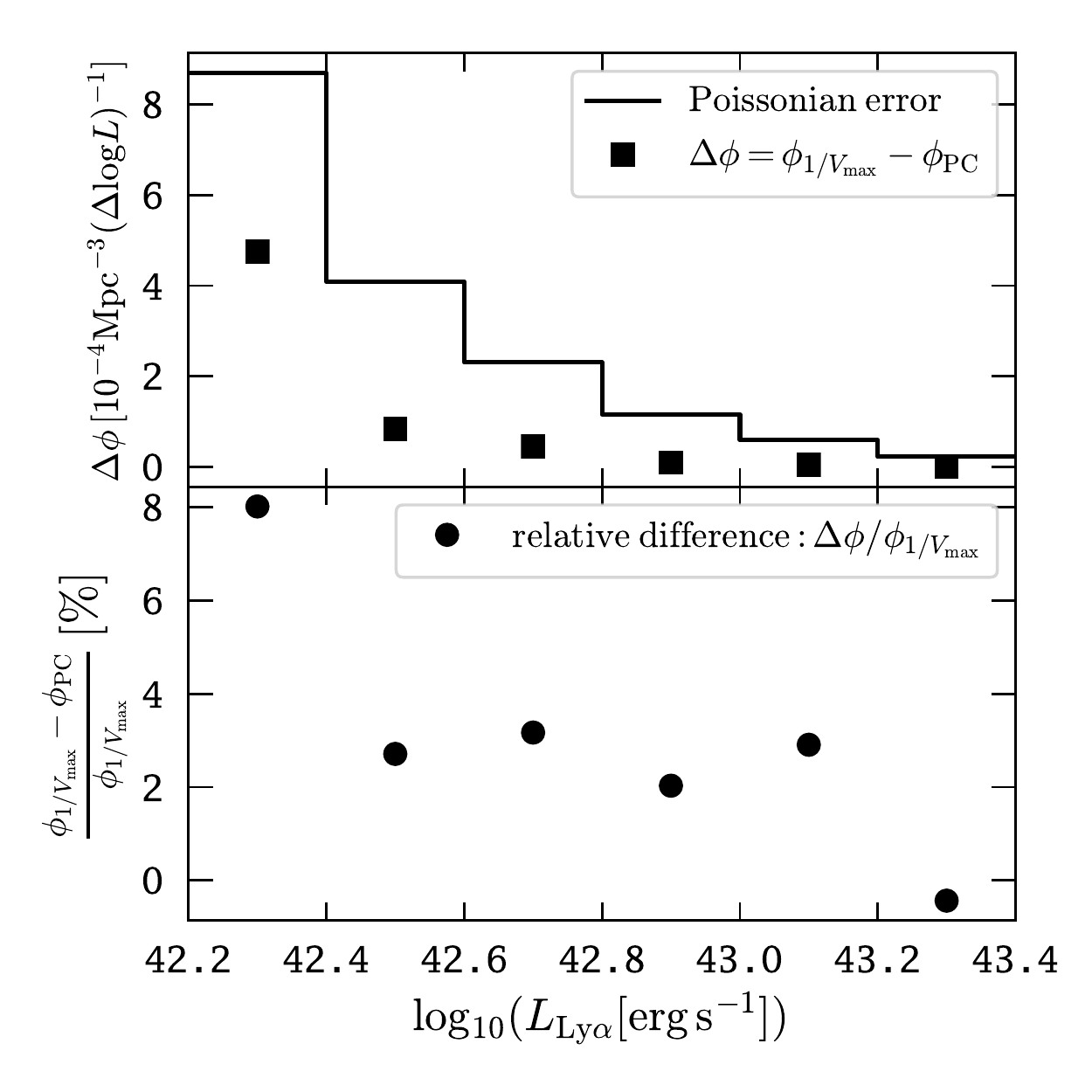}\vspace{-1em}
  \caption{\emph{Top panel}: Absolute difference between binned
    $1/V_\mathrm{max}$ and $\phi_\mathrm{PC}$ estimator for global MW
    LAE LF in comparison to the Poissonian errors in each
    bin. \emph{Bottom panel:} Relative difference (\%) between the two
    binned estimators.}
  \label{fig:phiestvmaxcomp}
\end{figure}

We first employed the non-parametric statistical test described in
Sect.~\ref{sec:non-parametric-test} to investigate whether the
observed MW LAE LF is consistent with a pure density evolution
scenario.  Table~\ref{tab:taus} lists the obtained $\tau$-values from
Eq.~(\ref{eq:14}) along with the corresponding $p$-values under the
normal distribution approximation.  We calculated $\tau$ both for the
RSSF and the PSFF.  Moreover, we  tested evolution not only for the
full MW redshift range, but also within three redshift ranges:
$2.9 < z \leq 4$, $4 < z \leq 5$, and $5 < z \leq 6.7$.  Regardless of
the adopted selection function, we find that the pure density
evolution scenario cannot be rejected over the full redshift range
(i.e. $|\tau| < 1$, thus $p_0 > 0.16$), or over the three redshift
ranges. This means that over the dynamic range of probed Ly$\alpha$
luminosities the shape of the observed LAE LF remains unchanged at
$3 \lesssim z \lesssim 5$.  The test, however, is not sensitive for a
possible change in the normalisation.  However, we  demonstrate below
(see especially Figure~\ref{fig:dfs}) that such a change in
normalisation is also not required for the observed LAE LF.

A non-evolving apparent LAE LF is consistent with the result from the
NB imaging survey by \cite{Ouchi2008}.  This study found no
significant differences between the apparent LAE LF (i.e. uncorrected for
Ly$\alpha$ absorption by the intergalactic medium) in their
three surveyed redshift slices ($z\simeq \{3.1, 3.7, 5.7\}$).  On the
other hand, at first our result appears to be in tension with the
recently reported LAE LF evolution from $z\simeq 2.5$ to $z\simeq 6$
within the SC4K survey \citep{Sobral2017}, but the change in the
SC4K LAE LFs is driven by a decreasing number density of the highest
luminosity LAEs ($\log L_\mathrm{Ly\alpha} [\mathrm{erg\,s}^{-1}] \gtrsim 43.0$).
Unfortunately, with the current MW data we do not sample a large
enough number of  luminous LAEs to obtain a statistically robust
confirmation of this result.  Moreover, the current MW sample is also
not well populated with $z\gtrsim 5.5$ LAEs.  Thus, to date, we also cannot add constraints to the ongoing debate in the literature regarding
a possible LAE LF evolution between $z=5.7$ and $z=6.6$
\citep[][]{Ouchi2010,Santos2016,Konno2018}.

We now analyse the differences in the resulting LAE LF when employing
the two different selection functions constructed in
Sect.~\ref{sec:constr-muse-wide}.  To this end we plot in
Figure~\ref{fig:rssfpssfcomp} the resulting cumulative LAE LFs
obtained with the $C^{-}$ method (Sect.~\ref{sec:c--method}) for the
RSSF, which explicitly accounts for the extended low surface brightness
halos of LAEs (left panel in Figure~\ref{fig:selfun}), and for the
PSSF, which assumes LAEs to be compact PSF broadened sources (right panel
in Figure~\ref{fig:selfun} and \ref{fig:selfun_zlum}).  We find that
at the faint end of our probed luminosity range
($\log L_\mathrm{Ly\alpha} [\mathrm{erg\,s}^{-1}] = 42.2$) the inferred LAE density utilising
the RSSF is a factor of 2.5 higher compared to the PSSF:
$\Phi_\mathrm{RSSF}(\log L_\mathrm{Ly\alpha} [\mathrm{erg\,s}^{-1}]= 42.2) =
2\times10^{-3}$\,Mpc$^{-3}$, while
$\Phi_\mathrm{PSSF}(\log L_\mathrm{Ly\alpha} [\mathrm{erg\,s}^{-1}] = 42.2) =
8\times10^{-4}$\,Mpc$^{-3}$.

We argue that due to the ubiquity of extended Ly$\alpha$ emission
around LAEs, the RSSF represents a more realistic selection function.
Hence, we regard the LAE LF constructed with this completeness
correction as unbiased.  Since previous LAE LF determinations, except
\cite{Drake2017a}, have not accounted for an extended nature of LAEs in
their selection functions, we expect similar biases in their inferred
number densities close to their limiting luminosities.  We
 demonstrate in Sect.~\ref{sec:comp-with-liter} that our
PSSF completeness-corrected Ly$\alpha$ LF agrees better with most
literature estimates.  Therefore, we note that our PSSF LAE LF
estimates here only serve demonstrative purposes, while the RSSF
corrected estimate can be regarded as our best estimate.

Numerically, we obtain the same difference between the LAE LFs from
the different selection functions when utilising the
$1/V_\mathrm{max}$ estimator (Sect.~\ref{sec:1v_mathrmmax-method}).
To demonstrate the similarity in the resulting LFs between $C^{-}$ and
$1/V_\mathrm{max}$ we compare in Figure~\ref{fig:globla_comp} the
inferred cumulative LAE LFs from the two estimators.  The maximum
discrepancy occurs at the faint end of our probed luminosity range.
Here $1/V_\mathrm{max}$ provides slightly higher LAE densities than
$C^{-}$:
$\Phi_{1/V_\mathrm{max}}(\log L_\mathrm{Ly\alpha}
[\mathrm{erg\,s}^{-1}]= 42.2) = 1.2 \times \Phi_{C^{-}}(\log
L_\mathrm{Ly\alpha} [\mathrm{erg\,s}^{-1}] = 42.2)$.  The same result
is obtained for the PSSF.  As outlined in Sect.~\ref{sec:c--method},
while the $C^{-}$ construction requires only an evaluation of the
selection function at redshifts where objects are detected, the
$1/V_\mathrm{max}$ estimate requires the evaluation of an integral
over the selection function at all redshifts.  Since our selection
function stems from an extrapolation of the results from a source
insertion and recovery experiment at five discrete wavelengths, the
two estimators deal differently with possible uncertainties from this
extrapolation approach.  Encouragingly, the differences in the final
LAE LF result are small.  This validates the robustness of our
selection function construction.

Lastly, we compute binned estimates from our sample using the
bins motivated in Sect.~\ref{sec:binn-trunc-sample} with the
$1/V_\mathrm{max}$ (Sect.~\ref{sec:1v_mathrmmax-method}) and
$\phi_\mathrm{PC}$ (Sect.~\ref{sec:phi_m-meth-citepp}) estimators. The
results are given in Table~\ref{tab:bin_lf}.  In
Figure~\ref{fig:phiestvmaxcomp} we compare the results from the two
different estimators.  Following the expectation of \cite{Page2000},
the binned $1/V_\mathrm{max}$ estimator is biased to higher values of
the differential LF, especially in the low-luminosity bins near the completeness limit.
We find the maximum discrepancy in the lowest luminosity bin to be
8\%.  However, at the current size of the MW sample the results are
within the statistical counting error for each bin.  Nevertheless, we
encourage the use of the \cite{Page2000} estimator in future
constructions of the binned LAE LF with larger samples since it is
less biased compared to the classical $1/V_\mathrm{max}$ techniques in
the lowest luminosity bins of the sample \citep[see also][]{Yuan2013}.

\subsection{Parametric modelling}
\label{sec:parametric-results}

\begin{figure}
  \centering
  \includegraphics[width=0.5\textwidth]{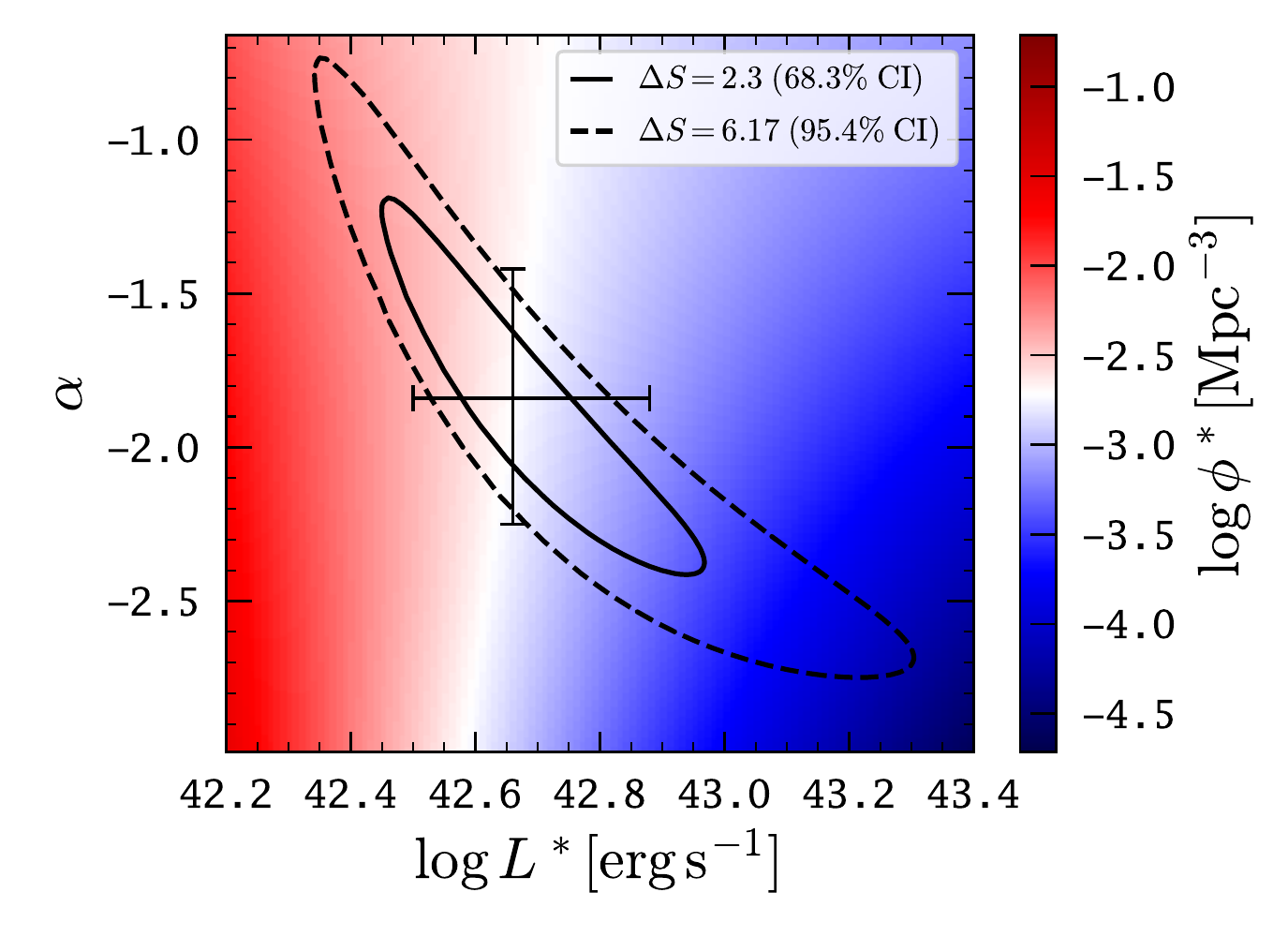}\vspace{-1em}
  \caption{Results from the Schechter function ML fit for the global
    MW LAE LF.  Contours are drawn at $\Delta S = \{2.3,6.17\}$
    thereby outlining the $\{68.3\%,95.4\%\}$ confidence intervals for
    $\alpha$ and $\log\,L^*$.  In colour we show the normalisation
    $\log\,\phi^*$, which is a dependent quantity on $\alpha$ and
    $L^*$, i.e. it is not a free parameter in the fitting procedure.
    The cross indicates the best-fitting
    $(\log\,L^* [\mathrm{erg\,s}^{-1}],\alpha)=(42.66,-1.84)$. At this
    point in $\log\,L^* - \alpha$ space the dependent normalisation is
    $\log \phi^*(\log\,L^*,\alpha) [\mathrm{Mpc}^{-3}]= -2.71$.  The
    1D error bars show the $68.3\%$ confidence interval from the
    marginalised distribution in $\alpha$ and $\log\,L^*$ (see text).}
  \label{fig:smap_plotter}
\end{figure}

\begin{figure}
  \centering
  \includegraphics[width=0.46\textwidth]{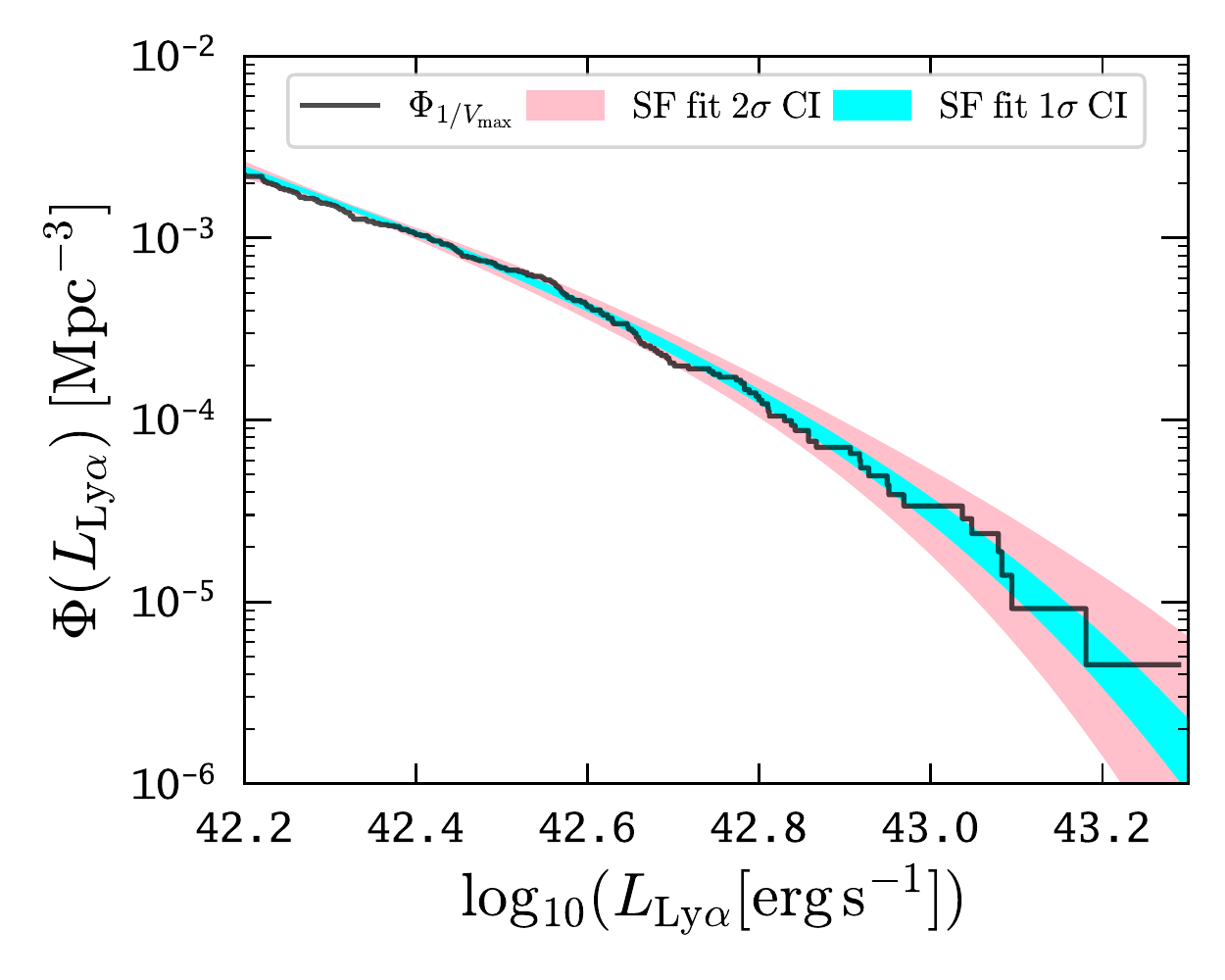}
  \vspace{-1em}
  \caption{Cumulative LAE LF from MW obtained with the
    $1/V_\mathrm{max}$ estimator in comparison to 68.3\% and 95.4\%
    confidence limits of the ML Schechter fit.}
\label{fig:cum_lum_plotter_with_fit}
\end{figure}

\begin{figure*}
  \centering
  \includegraphics[width=0.98\textwidth,trim=0 0 0 3,clip=true]{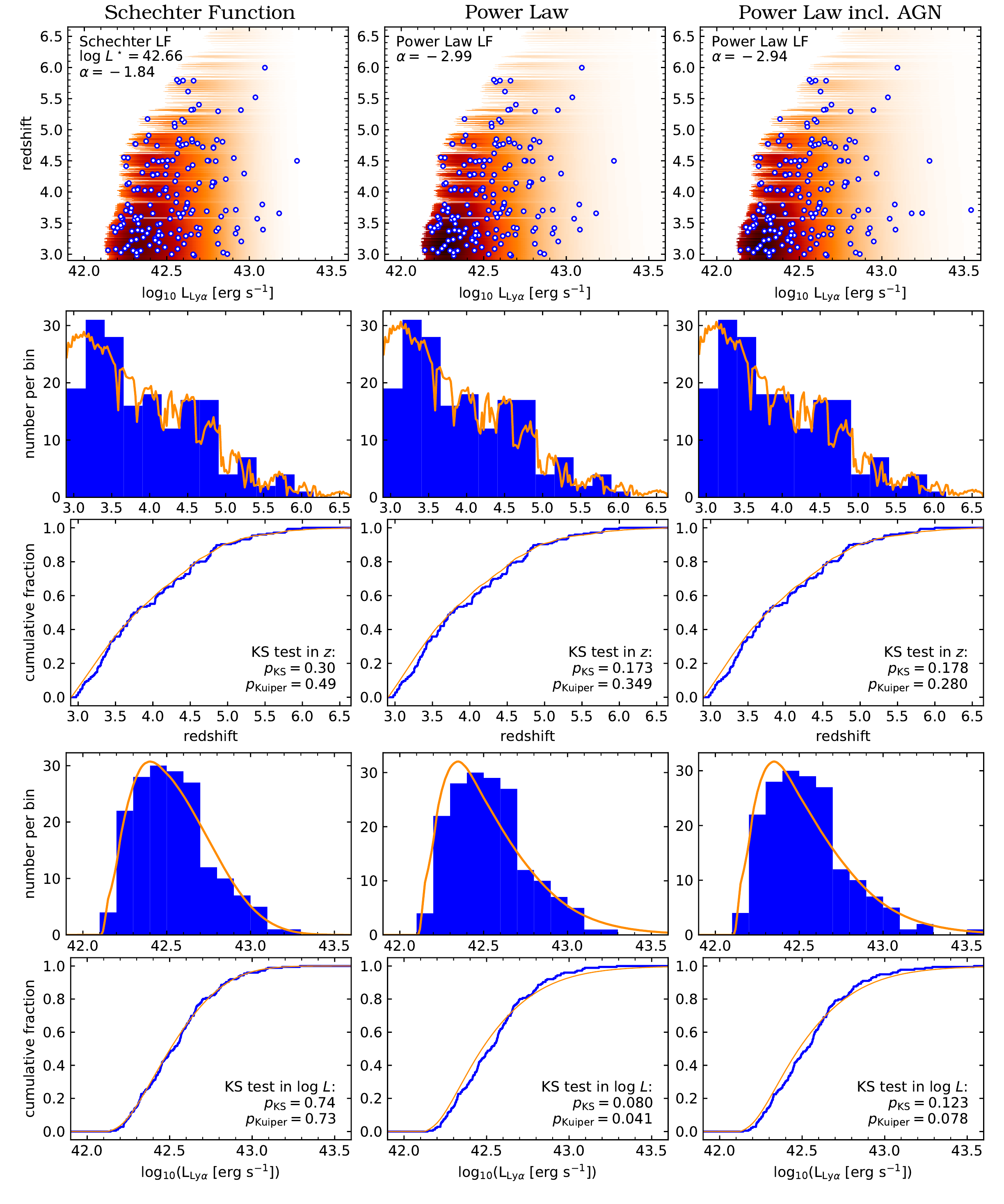} \vspace{-1.3em}
  \caption{Visualisation of the procedure to calculate the Kuiper- and
    KS-test statistics for the Schechter model (\emph{left panels}),
    power-law model (\emph{centre panels}), and the power-law model
    fit to the LAE sample where the X-Ray identified AGN were not
    excluded (\emph{right panels}). The panels in the \emph{first row}
    show the expected LAE distribution in redshift-luminosity space
    from the best-fit parameterisations folded with the MW survey
    area, selection function, and $f_c=0.15$ truncation criterion in
    shaded orange, while blue circles show the actual LAE samples in
    redshift-luminosity space.  The 2D KS-test statistic is computed
    by comparing the actual samples to the model distributions.  The
    panels in the \emph{second row} show predicted number counts
    computed from the distributions in the first row as a function of
    redshift in comparison to a histogram of the observed number
    counts (blue histogram). In the \emph{third row} we show the
    comparison between the normalised cumulative distribution in
    redshift (blue line) and the cumulative distribution from the
    model (orange line). These curves are used to calculate the Kuiper
    and 1D KS-test in redshift. The panels in the \emph{fourth and
      fifth row} are similar to those in the  second and third row,
    respectively, but compare the differential and cumulative
    distribution in $L_\mathrm{Ly\alpha}$-space to the model
    distributions.}
  \label{fig:kstest_all}
\end{figure*}

\begin{table*}
  \caption{$p$-values from Monte Carlo calibrated KS and Kuiper tests
    of the observed distribution against maximum-likelihood LF models
    obtained by folding the maximum-likelihood Schechter
    (Eq.~(\ref{eq:schechter}) or power-law (Eq.~\ref{eq:16})
    parameterisations with the MW survey area and LAE selection
    function (RSSF), as well as the $f_c=0.15$ truncation criterion.
    One-dimensional KS and Kuiper tests are performed in redshift and luminosities, while the 2D KS test operates directly
    in redshift-luminosity space (see Figure~\ref{fig:kstest_all}).  }
  \label{tab:kstests}
  \centering
  \begin{tabular}[]{lccccc}
    \hline \hline  \noalign{\smallskip}
    Parameterisation & $p_\mathrm{KS}^{L_\mathrm{Ly\alpha}}$ & $p_\mathrm{Kuiper}^{L_\mathrm{Ly\alpha}}$  & $p_\mathrm{KS}^z$ & $p_\mathrm{Kuiper}^z$ & $p_\mathrm{2DKS}$  \\  \hline  \noalign{\smallskip}
    Schechter  ($\log L^* = 42.66$, $\alpha=-1.84$,  $\log \phi^* = -2.71$)    & 0.74 & 0.73  & 0.30 & 0.49 & 0.87 \\ \noalign{\smallskip} 
    Power Law  ($\log L^* = 42.5$,  $\beta = -2.99$, $\log \phi^* = -2.932$)         & 0.08 & 0.04  & 0.17 & 0.35   & 0.23  \\ \noalign{\smallskip}
    Power Law incl. AGN ($\log L^* = 42.5$,  $\beta = -2.94$, $\log \phi^* = -2.930$) & 0.12 & 0.08  & 0.18  & 0.29  & 0.30 \\ \noalign{\smallskip} \hline \hline \noalign{\smallskip} 
  \end{tabular}
   \tablefoot{$L^*$ in $\mathrm{erg}\,s^{-1}$ and $\phi^*$ in Mpc$^{-3}$.  }
\end{table*}

\begin{figure*}
  \centering
  \includegraphics[width=\textwidth]{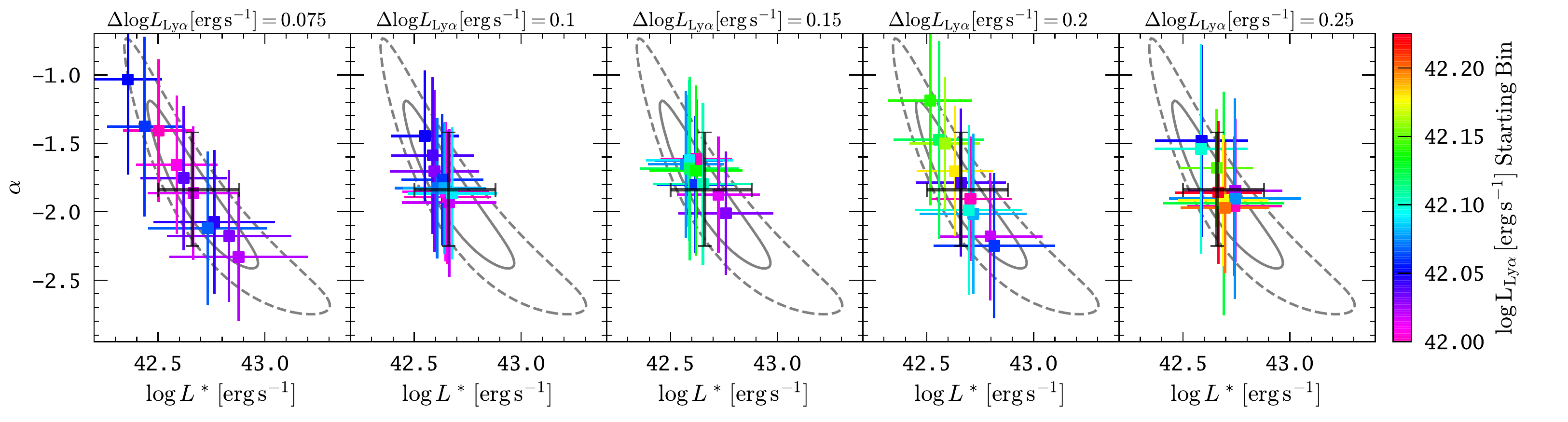}
  \caption{Resulting Schechter parameters $L^*$ and $\alpha$ from a
    non-linear least-squares fit (using the Levenberg-Marquardt
    algorithm) of the Schechter function (Eq.~\ref{eq:schechter}) to
    the binned differential $1/V_\mathrm{max}$ estimator
    (Eq.~\ref{eq:5}) for different binning schemes. Panels from left
    to right show five different bin sizes
    $\Delta \log L_\mathrm{Ly\alpha} [\mathrm{erg\,s}^{-1}] = \{
    0.075, 0.1, 0.15, 0.2, 0.25\}$.  Colour-coded in each panel is the
    best-fit  $(L^*,\alpha)$  pair according to the lowest-luminosity
    boundary of the starting bin. Incomplete bins (containing objects
    at $f_c < 0.15$) are ignored in the fit, i.e. the magenta point in
    the panel
    $\Delta \log L_\mathrm{Ly\alpha} [\mathrm{erg\,s}^{-1}] = 0.2$
    panel represents the adopted binning scheme
    (see Sect.~\ref{sec:binn-trunc-sample}) in Table~\ref{tab:bin_lf}
    and Figure~\ref{fig:comp1}.  For guidance the likelihood contours
    and the maximum-likelihood solution from
    Figure~\ref{fig:smap_plotter} are shown in each panel.}
  \label{fig:binfitvarbin}
\end{figure*}

\begin{figure*}
  \centering
  \includegraphics[width=\textwidth]{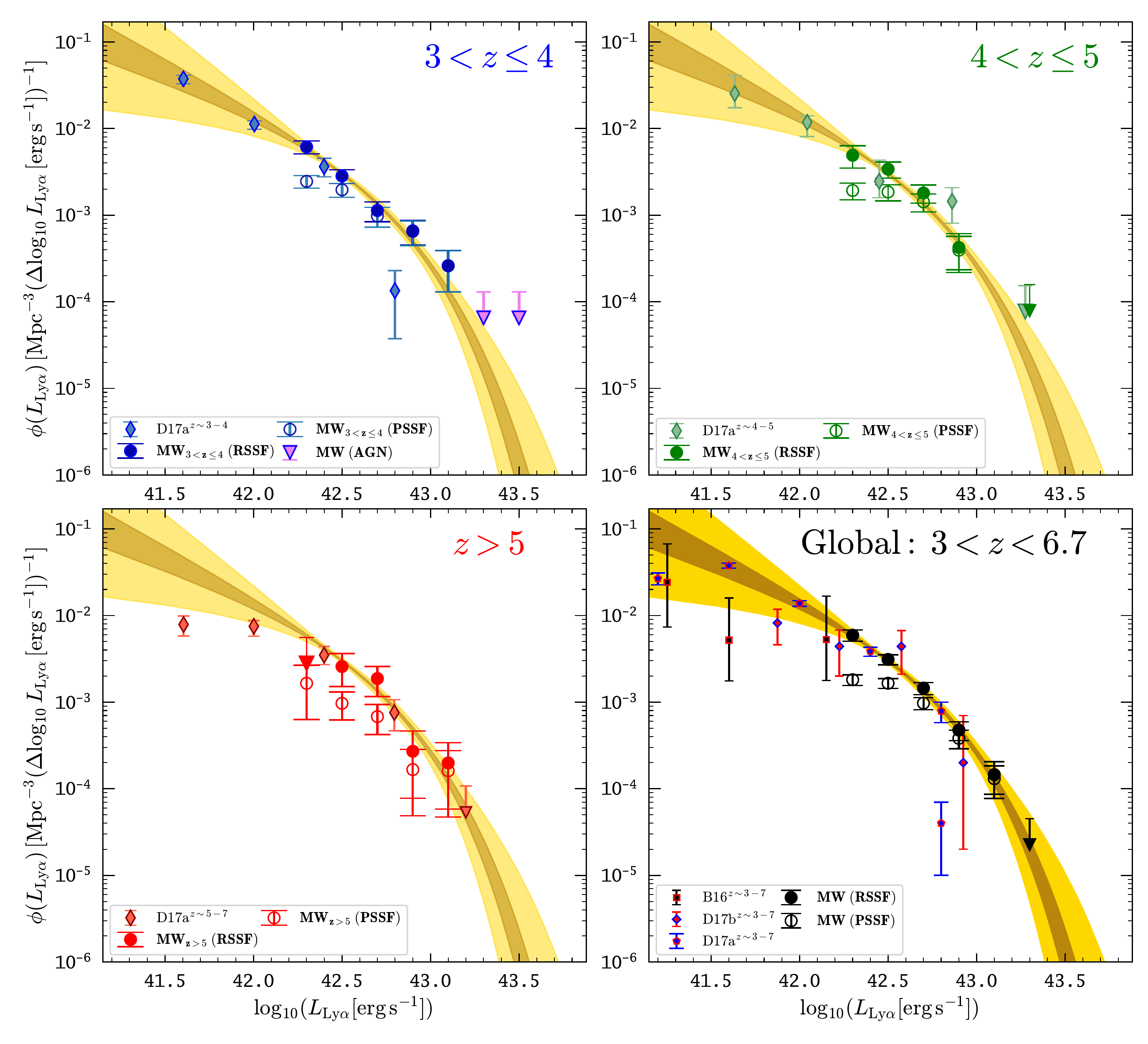}\vspace{-0.5em}
  \caption{Differential MUSE-Wide LAE LF in the redshift ranges
    $2.9 < z \leq 4$ (\emph{top left}), $4 < z \leq 5$
    (\emph{top right}), $5 < z < 6.7$ (\emph{bottom left}), and for the global MW redshift range (\emph{bottom
      right}).  Our RSSF (see Sect.~\ref{sec:select-funct-from-2}) corrected binned
    estimates are shown as filled circles, while  with the
    oversimplified PSSF (see
    Sect.~\ref{sec:select-funct-from-1}) corrected binned estimates
    are shown with open circles.  
    Yellow (dark yellow) shaded regions indicate the 68.3\%
    (95.4\%) confidence regions for a Schechter parameterisation
    obtained a maximum likelihood analysis
    (Sect.~\ref{sec:param-maxim-likel}).  For this parametric
    modelling we corrected with the RSSF for completeness.  Also shown
    in this figure are other MUSE LAE estimates obtained by the MUSE
    GTO consortium, namely the binned estimates by \cite{Drake2017}
    obtained from MUSE commissioning observations in the \emph{Hubble}
    Deep Field South, the binned estimates by \cite{Drake2017a} from
    the MUSE-Deep observations in the \emph{Hubble} Ultra Deep Field,
    and the pilot study by \cite{Bina2016} exploiting gravitational
    lensing by the lensing cluster Abell 1689. }
  \label{fig:dfs}
\end{figure*}

In order to obtain a parametric form of the LAE LF we evaluate the
inverted log-likelihood function in Eq.~(\ref{eq:17}) `brute-force'
for a densely sampled grid of the Schechter function
(Eq.~\ref{eq:schechter}) parameters $L^*$ and $\alpha$.  The minimum
of Eq.~(\ref{eq:17}) function represents the maximum-likelihood
solution.  It is found for
$\log L^* [\mathrm{erg\,s}^{-1}] = 42.66$ and $\alpha=-1.84$.  The
corresponding value for the normalisation $\phi^*$ is
$\log \phi^* [\mathrm{Mpc}^{-3}] = -2.71$.  In
Figure~\ref{fig:smap_plotter} the $\Delta S = 2.3$, and
$\Delta S = 6.17$ contours from the evaluation of Eq.~(\ref{eq:17})
are shown.  These two contours correspond to the standard $1\sigma$
and $2\sigma$ confidence regions (68.3\% and 95.4\%).  In this figure
we also visualise the dependence of the normalisation $\phi^*$ on
$L^*$ and $\alpha$.

From the `banana-shaped' appearance of the $\Delta S$ contours in
Figure~\ref{fig:smap_plotter} it is evident that we have a strong
degeneracy between $L^*$ and $\alpha$: higher $L^*$ values require
steeper faint-end slopes, i.e. smaller $\alpha$ values, and vice-versa.  By
marginalising over $\alpha$ and $L^*$ we recover the 1D
68.3\% confidence intervals
$\log L^* [\mathrm{erg\,s}^{-1}] = 42.66^{+0.22}_{-0.16}$ and
$\alpha = -1.84^{+0.42}_{-0.41}$, respectively.  These 1D
errors are also drawn as error bars around the maximum-likelihood
value in Figure~\ref{fig:smap_plotter}.  These Schechter parameters
are within the 68.3\% confidence intervals from the ML analysis
performed by \cite{Drake2017a} on the MUSE HUDF data:
$\log L^* [\mathrm{erg\,s}^{-1}] = \{42.72^{+0.23}_{-0.97},
42.74^{+\infty}_{-0.19}, 42.66^{+\infty}_{-0.19}\}$ and
$\alpha=\{ -2.03^{+0.76}_{-0.07}, -2.36^{+0.17}_{-\infty},
-2.86^{+0.76}_{-\infty}$ for the redshift ranges $2.9 \leq z \leq 4$,
$4 < z \leq 5$, and $5 < z \leq 6.64$, respectively.  We note that
in \cite{Drake2017a} the 1D confidence intervals on $L^*$ and $\alpha$
were estimated by taking the extremes of the $\Delta S = 1$ contour,
i.e. without doing the marginalisation.  This estimation implicitly
assumes a 2D Gaussian distribution for the likelihoods
\citep{James2006}.  Nevertheless, we verified that the extremes of the
$\Delta S = 1$ contour are in good agreement with the marginalised
confidence limits,  but we caution that such 1D errors do, by
construction, not reflect the interdependence between $\alpha$ and
$L^*$.  Importantly, this interdependence needs to be taken into
account when discussing the LAE LF redshift evolution based on
parametric LF fits\footnote{To facilitate this discussion in future
  work, we release our obtained $S(\log L_\mathrm{Ly\alpha}, \alpha)$
  and $\phi^*(\log L_\mathrm{Ly\alpha}, \alpha)$ functions shown in
  Figure~\ref{fig:smap_plotter} with this publication via the CDS.}.
 
In Figure~\ref{fig:cum_lum_plotter_with_fit} we compare the
maximum-likelihood estimated Schechter function LF to the
non-parametric $1/V_\mathrm{max}$ estimate.  The in this figure shown
68.3\% and 95.5\% confidence limits on the cumulative Schechter
function were obtained by randomly drawing\footnote{Random draws where
  realised with the rejection method \citep{Press1992}.} 1000 LAE LFs
from the normalised likelihood function (Eq.~\ref{eq:18}).  We
deliberately compare here the parametric results to the non-parametric
$1/V_\mathrm{max}$ estimate, because in both approaches the selection
function needs to be integrated over the whole redshift range (compare
denominators in Eq.~\ref{eq:19} and Eq.~\ref{eq:4}), which is not the
case for the $C^{-}$ method.  Thus, a comparison of the
maximum-likelihood results to the $C^{-}$ results would stand on
unequal footing.  As evident from
Figure~\ref{fig:cum_lum_plotter_with_fit}, there is excellent
agreement between the non-parametric and parametric LFs, indicating
that indeed the Schechter parameterisation appears qualitatively to be
a valid description of the LAE LF.  

We also test whether a  power law (Eq.~\ref{eq:16}) is a more
suitable parameterisation of the LAE LF from our MW data.
To this aim, we first calculate the inverted log-likelihood
function for a fine sampled grid of power-law slopes $\beta$.
We find the minimum in $S$ at $\beta = -2.99 \pm 0.12$.  The
normalisation, evaluated at $\log L^*=42.5$, is
$\log \phi^* = -2.932 \pm 0.006$.  We also perform the same analysis
without excluding the AGN from the sample.  In this case we recover a
slope $\beta = -2.94 \pm 0.12$ and normalisation
$\log \phi^* = -2.930 \pm 0.006$ (at $\log L^*=42.5$). 

Equipped with these results, we now quantify the goodness of fit. Our
statistical analysis  enables us to decide whether the power-law
or Schechter parameterisation describes the LAE LF more adequately.  A
possible statistical test in this respect is the Kuiper test
\citep[e.g.][]{Press1992,Ivezic2014}.  This test bears similarities to
the well-established Kolmogorov-Smirnov (KS) test, but it is more
sensitive to the discrepancies in the wings of the distribution
\citep[see also][]{Wisotzki1998}.  Hence, it is more suitable for the
situation at hand as the exponential cut-off to the power law in the
Schechter function modulates the expected frequency of the brighter
galaxies in our probed luminosity range.   Nevertheless, for comparison
purposes we also compute the classical KS tests.  Both tests are
one-dimensional, thus require marginalisation over our sample and the
model distributions (explained below), either over redshifts or
luminosities.  When marginalising over redshifts we thus test for
discrepancies between the observed and model luminosity distributions.
Given the assumption of a non-evolving LF over the probed redshifts,
which was already backed with evidence in the previous section, this
marginalisation provides the most powerful metric for testing the
different LF parameterisations.  Marginalising over luminosities, on
the other hand, tests whether the observed distributions in redshifts
are adequately described by one parameterisation.  This provides us
with a parametric test for redshift evolution.  Finally, dealing with
a 2D distribution in redshift-luminosity space we also calculate a 2D
variant of the KS statistic that was originally developed by
\cite{Peacock1983}.

A possible pitfall when utilising these tests is that we determined
the model parameters from the same dataset.  As explained in
\cite{Wisotzki1998},  the distribution functions of these test
statistics are thus not valid anymore for calculating the $p$-values needed to
reject or accept the null hypothesis that the  `data is represented by the
model'. This is because the null hypothesis model has been moved
closer to the data due to its estimation from the data.  We circumvent
this by performing Monte Carlo simulations to calculate the
distribution of these test statistics under the null hypothesis
\citep[][Chapt. 14.3]{Press1992}.  Therefore, we draw a large number of
samples of the same size as our LAE LF sample from the ML luminosity
function models.  In these simulations we account for the surveyed
area, the MW selection function (RSSF), and the $f_c=0.15$ sample
truncation criterion.  We show in Figure~\ref{fig:kstest_all}
the resulting 2D distributions in redshift-luminosity space from the
maximum-likelihood models together with the MW LAE sample.  Moreover,
we also show in this figure the marginalised differential and
cumulative distributions in redshift- or luminosity-space, together
with binned histograms of the actual samples for the Schechter
function, for the power law, and for the power law without exclusion
of the two AGN in the sample.  The 2D KS test-statistics are computed
by comparing the 2D model distributions to the actual samples, and the
1D KS- and Kuiper-tests are computed by comparing the cumulative 1D
model distributions to the cumulative sample distribution.  We list in
Table~\ref{tab:kstests} the resulting $p$-values from those tests.

It is already visually  apparent, especially when contrasting the
panels comparing the cumulative distributions in $L_\mathrm{Ly\alpha}$
in Figure~\ref{fig:kstest_all}, that the expected
distributions from the power-law parameterisations show marked
discrepancies with respect to the observed distribution.  This visual
impression is confirmed by the $p$-values (Table~\ref{tab:kstests}).
All three statistical tests result in markedly smaller $p$-values for the
power-law model compared to those for the Schechter model.  The
KS- and Kuiper-tests in redshift space result in $p$-values for which
neither the power-law model nor the Schechter model can be formally rejected.
This shows that a single parameterisation of the LAE LF is adequate to
describe the LAE LF over the redshift and luminosity range probed by
MW, but the Schechter model can be favoured due to its markedly higher
$p$-values.  This result is consistent with the non-parametric test
presented in the previous section that indicated a non-evolving LAE LF
over the redshift range probed by MW (Table~\ref{tab:taus}).  Given
the non-evolving LF, the resulting $p=0.04$ of the Kuiper test in
$L_\mathrm{Ly\alpha}$ for the power-law model means that we can reject
this parameterisation at $2\sigma$ significance.  Only when not
excluding the X-ray identified AGN from the LAE sample, the power law
becomes a marginally consistent description of the sample.  Based on
these results we adopt our ML Schechter model as the working
hypothesis for the remainder of this paper.

We note that parametric models for LAE LFs in the literature are
sometimes obtained by $\chi^2$ fitting a model to non-parametric
binned estimates of the differential LF
\citep[e.g.][]{vanBreukelen2005,Cassata2011,Matthee2015,Santos2016}.
However, we caution that the resulting model parameters and
  uncertainties depend on the placement and width of the bins. We
visualise this  for our sample in Figure~\ref{fig:binfitvarbin}. There
we show the resulting $(L^*,\alpha)$ values from a non-linear fit
(obtained with the Leveneberg-Marquardt algorithm) of the Schechter
function (Eq.~\ref{eq:schechter}) to different binned
$1/V_\mathrm{max}$ estimates (Eq.~\ref{eq:5}). For this exercise we
varied both the size (different panels in
Figure~\ref{fig:binfitvarbin}) and the placement (different colours in
Figure~\ref{fig:binfitvarbin}) of the bins.  Moreover, we ignored
incomplete bins, i.e. bins with objects that fall below the $f_c=0.15$
truncation criterion, in the fitting procedure.  As is evident, the
resulting $(L^*,\alpha)$ pairs  scatter substantially, with only a few
combinations of bin-width and bin-placement reproducing the actual ML
solution.  Thus, this fitting approach will not lead to a robust
parameterisation of the LF.  However, given a ML solution it could
potentially be used to determine an optimal bin-width and
bin-placement at which the binned estimate will be closest to the
adopted parametric form.  Indeed, for our adopted bin-width
($\Delta \log L_\mathrm{Ly\alpha} [\mathrm{erg\,s}^{-1}] = 0.2$) and
bin-placement (lowest luminosity boundary
$\log L_\mathrm{Ly\alpha} [\mathrm{erg\,s}^{-1}] = 42.2$), the
parametric fit to the binned data is in very good agreement with the
ML solution.

We plot in Figure~\ref{fig:dfs} the non-parametric differential MW LAE
LF in three redshift bins ($2.9 < z \leq 4$, $4 < z \leq 5$, and
$5 < z \leq 6.7$), as well as the global ($2.9 < z < 6.7$) LAE LF.
The non-parametric results shown in this figure are obtained with the
$\phi_\mathrm{PC}$ method for the RSSF and the PSSF.  For both the
redshift bins and the whole redshift range we also display the 68.3\%
and 95.4\% confidence intervals of the global Schechter LF.  As for
Figure~\ref{fig:cum_lum_plotter_with_fit}, these intervals were
obtained by randomly drawing 1000 LAE LFs from the normalised
likelihood function.  Here it can be seen that the global Schechter
fit is an excellent description of the global binned RSSF LF.  This
result confirms what we saw already when comparing the parametric to
the non-parametric cumulative LAE LFs in
Figure~\ref{fig:cum_lum_plotter_with_fit}.  Moreover, the binned
estimates in the different redshift bins are also in excellent
agreement with the global Schechter parameterisation, thus adding
further evidence to our previous tests that indicated a non-evolving
apparent LAE LF.  All these results justify the use of a global LAE LF
in this redshift range by MW.  Hence, the estimates in the redshift
bins here serve only demonstrative purposes and will not be considered
further.  For the same reason, parametric estimates in the redshift
bins are prohibitive for our sample as they just would lead to a
larger uncertainty on the final fitting parameters (known as 
overfitting).  We have commented already on the upward correction of
the LAE LF by up to a factor of 2.5 at the faint end  of our probed
luminosity range when utilising the RSSF instead of the PSSF
(Sect.~\ref{sec:non-param-results}).  Finally, in the  
comparison between RSSF and PSSF corrected LFs presented here, it can be seen that neglecting extended Ly$\alpha$ emission in the
selection function naturally leads to the inference of a flatter
faint-end slope $\alpha$ in the Schechter parameterisation.  We 
demonstrate in the next section that the PSSF corrected values are in
better agreement with previously determined literature estimates.

We also compare in Figure~\ref{fig:dfs} the MW LAE LF to published LAE
LF estimates  from other MUSE surveys performed within the MUSE
consortium \citep[][]{Bina2016,Drake2017,Drake2017a}. Key parameters
from these surveys are also listed in Table~\ref{tab:litcomp}. Both
the binned estimates from the pilot study by \cite{Bina2016}, which
makes use of the lensing cluster Abell 1689, as well as the global LAE
LF determination from the deep MUSE commissioning data in the
\emph{Hubble} Deep Field South show some agreement at the 1$\sigma$
level with our estimates.  However, the error bars from these early
analyses of MUSE data are quite large, and the estimates scatter
substantially.  More relevant is the good agreement between our
results and the binned estimates from the MUSE-Deep programme in the
\emph{Hubble} Ultra Deep Field by \cite{Drake2017a}.  Where the
luminosity ranges between MUSE-Deep and the  MW sample presented here
overlap, the data points are in almost perfect agreement, except for
the brightest \cite{Drake2017a} bins for the redshift range
$2.9 < z \leq 4$ and for the global LF.  However, the mismatch in
those brightest bins is a consequence of the pencil-beam nature of the
MUSE Deep survey, making it prone to cosmic variance for such brighter
and rarer LAEs.  We  again note that the \cite{Drake2017a} study also
incorporates a correction for extended Ly$\alpha$ halos in their
completeness function estimates.  In this respect it is especially
encouraging that even their faintest bins
($\log L_\mathrm{Ly\alpha} [\mathrm{erg\,s}^{-1}] < 42.0$) are in
agreement with the $2\sigma$ contours of our extrapolated Schechter
parameterisation below the luminosity limit of MW\footnote{As
    discussed in \cite{Drake2017a}, their faintest bins at
    $3 < z < 4 $ are consistent with the LAE LF construction at
    $z\sim3$ from a blind 92h long-slit integration with FORS2 by
    \cite{Rauch2008}.}, except at $z>5$.  There, however, the
faintest bins are below the adopted $f_c = 0.25$ completeness cut-off
for the parametric modelling in \cite{Drake2017a} as at these low
completeness levels the selection function was deemed unreliable.  The
comparison with the MUSE-deep analyses demonstrates how MW is
complementary at brighter luminosities.  In a forthcoming study we
will perform a joint and homogenised LAE LF analysis of the deep and
wide MUSE datasets.

\section{Comparison with the literature}
\label{sec:comp-with-liter}

\begin{table*}
  \caption{Compilation of key parameters of LAE LF studies from the
    literature in the redshift range probed by MW.}
  \label{tab:litcomp}
  \centering
  \begin{tabular}{lcccccccc}
    \hline \hline  \noalign{\smallskip}
    Reference        & Method$^{(*)}$ & $\mathrm{EW}_\mathrm{Ly\alpha}^{\mathrm{lim}}$  & $z$          & Area      & Volume                 & $\log L_\mathrm{Ly\alpha,lim}$ & $N_\mathrm{phot}^{(**)}$ & $N_\mathrm{spec}^{(**)}$ \\
        {}           &   {}   &    [\AA{}]                         &  {}          & [deg$^2$] & [Mpc$^3$]                & $\log$ [erg\,s$^{-1}$]          &     {}           &  {}     \\
    \noalign{\smallskip}\hline \noalign{\smallskip}
    \cite{Ouchi2008} & SC/NB503 & 64                               & $3.1\pm0.03$  & 0.983     & $7 \times 10^5$  & 42.0                           & 356               & 41 \\
    $\dots$          & SC/NB570 & 44                               & $3.7\pm0.03$  & 0.965     & $6.1 \times 10^5$& 42.6                           & 101               & 26 \\
    $\dots$          & SC/NB816 & 27                               & $5.7\pm0.05$  & 1.033     &$9.2 \times 10^5$ & 42.5                           & 401               & 17 \\ \hdashline     \noalign{\smallskip}
    \cite{Grove2009} & FORS2/3NB & 25                              & 2.85/3.15/3.25& 0.037     &$1.4 \times 10^4$ & 41.5                          & 83                & 59 \\ \hdashline     \noalign{\smallskip}
    \cite{Cassata2011} & VIMOS   & -                               & $2 - 6.6$ & 0.62/0.16$^\dagger$ & -               & 41.0                           & 84$^{\dagger\dagger}$ & 153$^{\dagger\dagger}$ \\ \hdashline \noalign{\smallskip}
    \cite{Matthee2017} & INT/NB501 & 12                         & $3.06 \leq z \leq 3.17$ & 0.7 & $7.2 \times 10^5$    & 43.0                           & 32                & 5 \\ \hdashline \noalign{\smallskip}
    \cite{Sobral2017} & \footnotesize{SC/IA464-527}$^\sharp$ & 50   & $3.1 \pm 0.4$ & $\approx 2$ &$17.3\times 10^6$          & 42.5                      & 2146               & ?$^\ddagger$ \\
    $\dots$           & \footnotesize{SC/IA574,624}         & 50   & $3.9 \pm 0.3$ & $\approx 2$ &$10.1\times 10^6$     & 42.95                          & 240                  & ?$^\ddagger$ \\
    $\dots$           & \footnotesize{SC/IA679,709}         & 50   & $4.7 \pm 0.2$ & $\approx 2$ &$10.6\times 10^6$    &  43.1                           & 160                 & ?$^\ddagger$ \\
    $\dots$           & \footnotesize{SC/IA738-827}$^\natural$&50   & $5.4 \pm 0.5$ & $\approx 2$ &$15.5\times 10^6$    & 43.3                            & 147                 & ?$^\ddagger$ \\
    $\dots$           & SC4K/global                 & 50   &  $\sim 2.5 - 6$& $\approx 2$&$\sim10^8$ &   42.5 & 3434 &   112$^\ddagger$ \\
    \hdashline \noalign{\smallskip}
    \cite{Drake2017a} & MUSE/HUDF  & - & $3.5\pm0.5$ & $7\times10^{-7}$ & $3.1\times10^{4}$ & 41.0  & - & 193 \\
    $\dots$           & $\dots$    & - & $4.5\pm0.5$ & $7\times10^{-7}$ & $2.6\times10^{4}$ & 41.0  & - & 144 \\
    $\dots$           & $\dots$    & - & $5.8\pm0.8$ & $7\times10^{-7}$ & $3.6\times10^{4}$ & 41.0  & - & 50 \\
    $\dots$           & $\dots$    & - &$2.91 - 6.64$& $7\times10^{-7}$ & $9.3\times10^{4}$ & 41.0  & - & 387 \\ \hdashline \noalign{\smallskip}
    \cite{Drake2017}  & MUSE/HDFS  & - &$2.91 - 6.64$& $8\times10^{-8}$ & $1\times10^{4}$   & 41.4  & - & 59 \\ \hdashline \noalign{\smallskip}
    \cite{Dawson2007} & 4m/5NB     & 15 & $4.4\pm0.1$ & $2\times10^{-4}$  & $1.5\times10^{4}$ & 42.0  & 97 & 79 \\ \hdashline \noalign{\smallskip}
    \cite{Shioya2009} & SC/NB711   & 12 &$4.86\pm0.03$& 1.83            &  $1.1\times10^6$    & 42.8  & 79 & 0  \\ \hdashline \noalign{\smallskip}
    \cite{Konno2018}  & HSC/NB816  & 10 &$5.73\pm0.05$& 13.8            & $1.2\times10^7$     & 42.9  & 1077 & 49 \\ \hdashline \noalign{\smallskip}
  \footnotesize{\cite{Shimasaku2006}}& SC/NB816   & 10 &$5.7\pm0.05$ & 0.2             & $1.8\times10^5$     & 42.5  & 89   & 39 \\ \hdashline \noalign{\smallskip}
    \cite{Santos2016}   & SC/NB816   & 25 &$5.7\pm0.05$ & 7               & $ 6 \times 10^6$    & 42.4  & 514   & 46 \\ \hdashline \noalign{\smallskip}
    \cite{Henry2012}   & \footnotesize{IMACS/NB+slits} & - & $5.7 \pm 0.1$& 0.015 & $1.5\times10^4$ & 42.1 & 105$^{\ddagger\ddagger}$ & 6 \\ \hdashline \noalign{\smallskip}
    \cite{Bina2016}    & \footnotesize{MUSE/Abell1689} & - & $2.91 - 6.64$  & $8\times10^{-8}$ & $900^\flat$& 40.5 & - & 17  \\
    \hline 
  \end{tabular}
  \tablefoot{\\
    $^{(*)}$:  Legend for abbreviations: SC/X$=$Subaru Suprime-Cam with filter X; HSC/NB816$=$Subaru Hyper Suprime-Cam with NB816 filter; FORS2/3NB$=$ESO VLT/FORS2 - 3 fields, with three different narrow-band filters; VIMOS$=$ESO VLT/VIMOS multi-slit spectroscopic survey; INT/NB501$=$  Wide-Field Camera with NB501 filter at Isaac Newton 2.5m Telescope; MUSE/HUDF$=$MUSE Hubble Ultra Deep Field; MUSE/HDFS$=$MUSE Hubble Deep Field South; 4m/5NB$=$5 overlapping narrow-band filters on two 4m class telescopes; IMACS/NB+slits$=$Multi-slit narrow-band spectroscopic search with IMACS on the Baade telescope \citep[see also][]{Martin2008}; MUSE/Abel1689$=$MUSE observations of the lensing cluster Abell\,1689. \\
    $^{(**)}:$ Number of photometrically selected LAE candidates ($N_\mathrm{phot}$) and number of spectroscopic confirmations ($N_\mathrm{spec}$). \\
    $^\dagger$: Area of the imaging campaign (VIMOS Deep Survey / VIMOS Ultra Deep Survey) from which targets were pre-selected for VIMOS multi-slit spectroscopy. \\
    $^{\dagger\dagger}$: Here $N_\mathrm{phot}$ refers to the number of photometrically pre-selected LAEs, while $N_\mathrm{spec}$ indicates the number of serendipitously detected sources. \\
    $^\ddagger$: Spectroscopic confirmations only reported for the combined SC4K sample. \\
    $^{\ddagger\ddagger}$: Single unresolved emission line objects without continuum detections \citep[see also][]{Martin2008}. \\
    $^\sharp$: IA464, IA484, IA505, and IA527 medium-band filters. \\
    $^\natural$: IA738, IA767, and IA827 medium-band filters. \\
    $^\flat$: Effective comoving volume from lensing magnification.  }
\end{table*}

\begin{figure}
  \centering
  \includegraphics[width=0.45\textwidth]{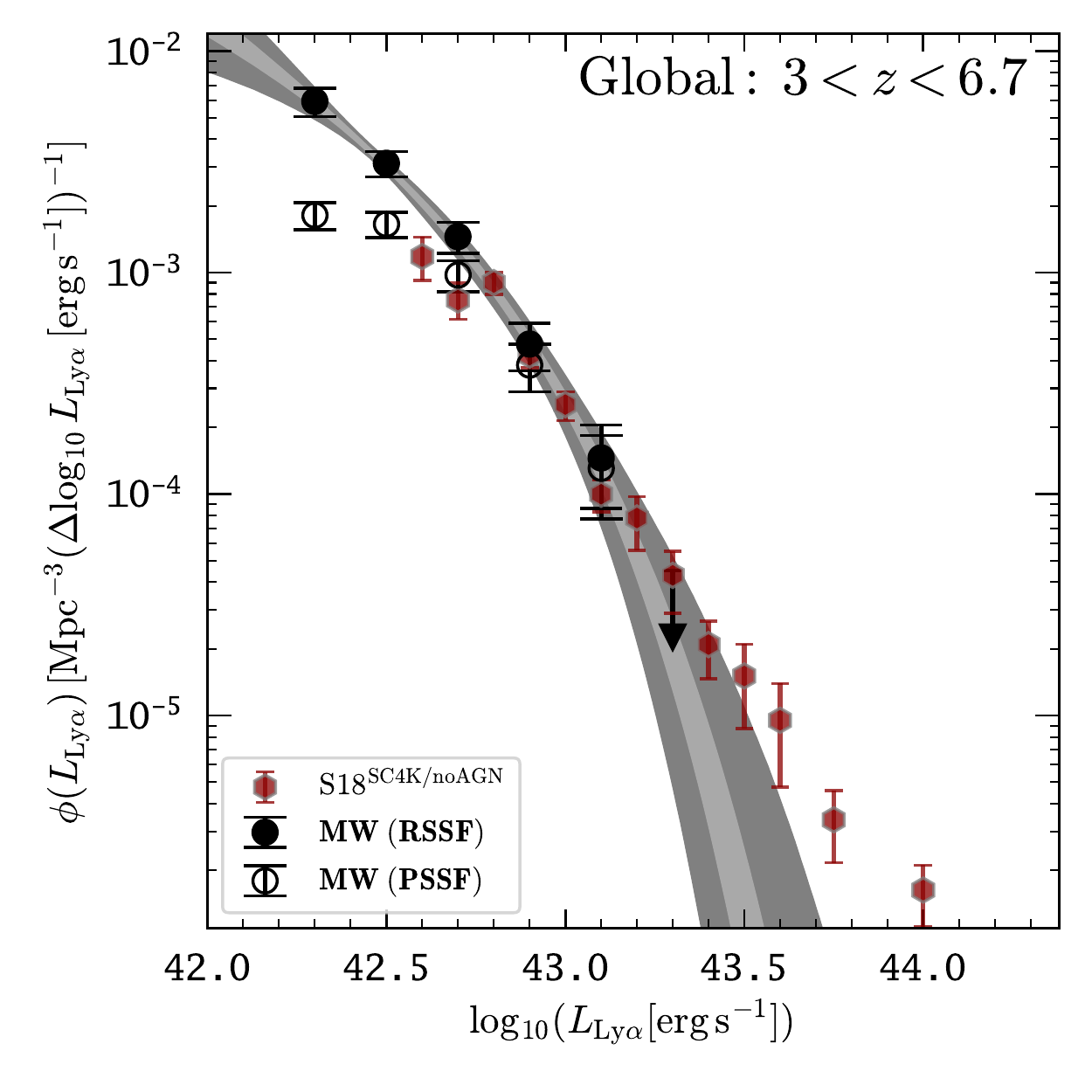}\vspace{-1em}
  \caption{Global ($2.9 \leq z \leq 6$) MUSE-Wide LAE LF (binned RSSF
    and PSSF corrected results from filled and open circles,
    respectively) with 1$\sigma$ (dark grey shaded region) and
    2$\sigma$ intervals (light grey shaded region) for the RSSF corrected Schechter
    parameterisation as in the bottom right panel of
    Figure~\ref{fig:dfs}, in comparison to the binned estimates of the
    global ($2.5 \leq z \leq 6$) SC4K LAE LF \citep{Sobral2017}.}
  \label{fig:comp1}
\end{figure}

\begin{figure}
  \centering
  \includegraphics[width=0.44\textwidth]{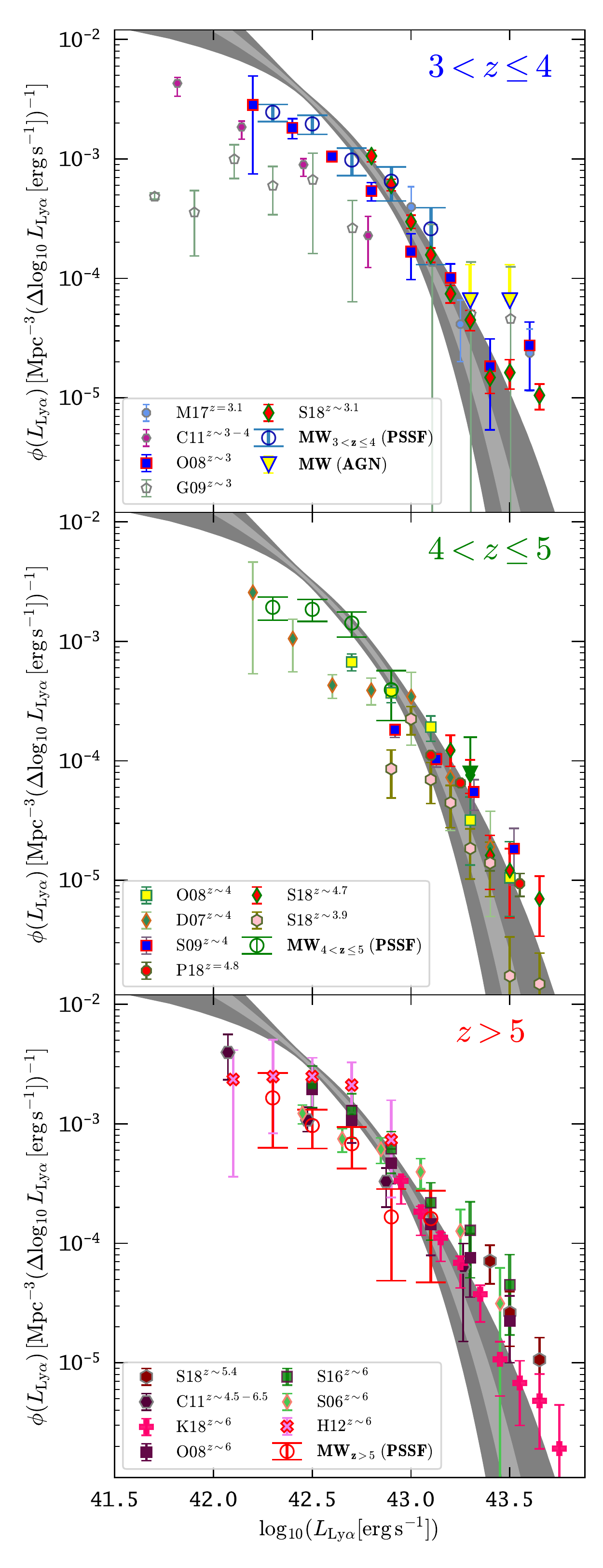}\vspace{-1.2em}
  \caption{Differential LAE LF estimates from the literature grouped
    in three redshift bins ($2.9 < z \leq 4$ in the \emph{top panel},
    $4 < z \leq 5$ in the \emph{middle panel}, and $5 < z \lesssim 6$
    in the \emph{bottom panel}) compared to our 1$\sigma$ (dark grey
    shaded region) and 2$\sigma$ intervals (light grey shaded region) for
    the RSSF corrected global Schechter parameterisation (shown
    already in Figure~\ref{fig:dfs} and Figure~\ref{fig:comp1}).
    References are provided in Table~\ref{tab:litcomp}; in the legend
    we abbreviate  with the first letter of the first author and
    the last two digits of publication year.  We also show our PSSF
    corrected binned estimates as open circles, as they are often in
    better agreement with the literature estimates (see text).}
  \label{fig:comp2}
\end{figure}

We now compare the obtained MW LAE LF with previous literature
estimates in the redshift range $3 \lesssim z \lesssim 6$.  For this
purpose we utilise the literature compilation of binned differential
LAE LF estimates provided by \cite{Sobral2017}, with the exception of
a few references which were not present in that compilation (namely
the studies by \citealt{Shimasaku2006}, \citealt{Shioya2009},
\citealt{Henry2012},\footnote{We use the `inferred
    LAEs, high LF' estimate from \cite{Henry2012}, for which  the
    less certain LAEs were also kept in the sample.} and
\citealt{Konno2018}).  An overview of the comparison studies is
provided in Table~\ref{tab:litcomp}, where we list their methodology,
redshift ranges, survey areas, probed comoving volumes, as well as the
lowest Ly$\alpha$ luminosities to which the LAE LF was probed.  For
the imaging campaigns we also list the adopted equivalent width cuts,
as well as the number of photometric LAE candidates and actual
spectroscopic confirmations.

Except for the MUSE studies mentioned at the end of the previous
section only \cite{Sobral2017} attempted to construct a global LAE LF
over a similar redshift range.  We provide a comparison between their
binned estimates and our binned and parametric estimates in
Figure~\ref{fig:comp1}.  Where the MW luminosity range overlaps with
SC4K, the two LF estimates are in agreement, except for the faintest SC4K
bins.  These bins fall below our RSSF corrected results and line up
more closely with our PSSF corrected binned estimates.  We  comment on
this mismatch at the faint end below, as it seems to be a generic
property of previous LAE LF construction attempts.

First we focus in Figure~\ref{fig:comp1} on the bright end of the
global SC4K LAE LF ($\log L_\mathrm{Ly\alpha} \gtrsim 43.2$). There we
note an apparent excess of the \cite{Sobral2017} bins compared to the
$1\sigma$ contours of our extrapolated Schechter parameterisation.  The display of our binned RSSF-corrected estimate together with
the SC4K binned estimate is in fact very suggestive of a non-existent `knee'
in the LAE LF and thus  supportive of the power-law
parameterisation favoured by \cite{Sobral2017}.  This is not in
tension with our statistical analysis presented in
Sect.~\ref{sec:parametric-results} that disfavoured a power law.  The
reason could simply be the limited dynamical range in high Ly$\alpha$
luminosities.  Such bright and rare LAEs are only sampled with robust
statistics in wide-field NB imaging campaigns.  Notably,  several
other studies also indicate that a non-exponential drop-off at the
bright end of the LAE LF is not required, both at lower redshifts
\citep[$z\lesssim 2$,][]{Konno2016,Wold2017,Hao2018} and at higher
redshifts \citep[$z\gtrsim 5$,][]{Santos2016,Matthee2017,Bagley2017}.
While the low redshift studies demonstrate convincingly that the
excess at the bright end of the LAE LF can almost exclusively be
attributed to AGN \citep[see especially][]{Konno2016,Wold2017}, the
nature of these sources at high redshifts is  less clear.
Another hint at the possible mismatch of our favoured Schechter model
with the bright end of the LAE LF can also be seen in
Figure~\ref{fig:comp2}, where we compare the 1$\sigma$ and 2$\sigma$
contours of the global Schechter parameterisation from our likelihood
analysis with binned estimates from the literature in different
redshift ranges. However, there is considerable scatter amongst the
literature estimates, even for the different redshift slices from
the SC4K survey, and at least most of the data points are consistent
at the 2$\sigma$ level with the Schechter model.

If the bright-end excess seen in the LAE LF cannot be attributed to
AGN activity (e.g. \citealt{Sobral2017} excluded AGN based on X-Ray
and radio diagnostics), then the LAE LF would have a different shape
compared to the rest-frame ultraviolet (UV) LF of high-redshift
galaxies which appears to be well described by a Schechter function
\citep[e.g.][]{Bouwens2007,Bouwens2015a}.  However, the most recent
wide area ground-based surveys start to question this result by
reporting a bright-end excess in the UV LF that cannot be solely
attributed to AGN activity and seems to deviate from a simple
Schechter parameterisation \citep{Ono2018,Viironen2018}.  Certainly,
Ly$\alpha$ radiative transfer is expected to modulate the Ly$\alpha$
output of a galaxy compared to its overall ionising photon production,
which as a good first-order approximation can be traced by its UV
luminosity \citep[e.g.][]{Bouwens2016a,Schaerer2016}.  In principle
the UV and LAE LFs can be linked to each other
\citep{Henry2012,Gronke2015a}.  However, in which way
radiative transfer processes or additional Ly$\alpha$ photon
production processes \citep[e.g. ionising photons from UV undetected
satellite galaxies or Ly$\alpha$ boosting from the UV background as
proposed in ][]{Mas-Ribas2016} could influence the bright end of the
LAE LF compared to the bright end of the UV LF remains currently
purely speculative.  A few of the most-luminous LAES at
$z\gtrsim 6$ have already received observational attention
\citep[e.g.][]{Ouchi2009,Lidman2012,Hu2016,Matthee2018}, with one
object being suggested to either host metal-free stars
\citep{Sobral2015} or a direct-collapse black hole
\citep{Pallottini2015}. At $z\sim2 - 3$ \cite{Sobral2018} presented
recently spectroscopic results on 20 bright LAEs
($\log L_\mathrm{Ly\alpha} \,[\mathrm{erg\,s^{-1}}]>42.7$).
Interestingly, these authors report a 60\% AGN fraction for such
luminous LAEs, which rises sharply to 100\% for
$\log L_\mathrm{Ly\alpha} \,[\mathrm{erg\,s^{-1}}]>43.3$.  This indeed
suggests that the observed deviations from a Schechter function at
bright luminosities are caused by sources whose Ly$\alpha$
emission is powered by non-thermal black hole accretion
processes rather than star formation.

We also find some notable overall disagreements between the literature
and our estimates in luminosity range where MW overlaps with other
surveys.  In the redshift range $2.9 < z \leq 4$ (top left panel in
Figure~\ref{fig:comp2}) we find that our LF is significantly higher
(up to an order of magnitude) than the LF estimates
obtained by \cite{Cassata2011}.  However, the \citeauthor{Cassata2011}
$z\sim 3 - 4$ LF is also significantly below most other literature
estimates and it is only consistent with the faint end
($\log L_\mathrm{Ly\alpha} [\mathrm{erg\,s}^{-1}] \leq 42.5$) of the
\cite{Grove2009} LF.  Moreover,  most of the LF bins from
\cite{Dawson2007} $z\sim4$ (centre panel in Figure~\ref{fig:comp2})
are significant below our inferred LF.

Finally, we find from the comparison in Figure~\ref{fig:comp2}, where
we group the literature results in three redshift bins that the
majority of literature LF estimates at luminosities
$\log L_\mathrm{Ly\alpha} \lesssim 42.5$ fall below our global
Schechter parameterisation.  We note again that this
parameterisation was obtained by implicitly correcting for extended
low surface brightness Ly$\alpha$ halos by utilising our RSSF.  In
this respect it is especially interesting that that the majority of
the literature estimates are often in nearly perfect agreement with
our PSSF completeness corrected LF estimates.  Especially the binned
estimates of \cite{Ouchi2008} at $z\sim3$, as well as the binned
estimate from \cite{Shimasaku2006} and \cite{Cassata2011} at $z\sim6$
line up perfectly with our PSSF corrected estimates.  Thus, we are
able to reproduce the results of previous campaigns by using a
completeness correction that is comparable to the ones applied in
those studies. 

Notably, almost all LAE LF estimates in the literature to date have  not
taken the extended nature of LAEs into account when constructing their
selection functions.  For example, \cite{Ouchi2008} populate their
NB imaging data with fake point sources, while
\cite{Hao2018}, at $z\sim2$, rescale the flux of stellar images in
their images.  A slightly different approach was used by
\cite{Konno2018} who utilise a S\`{e}rsic $n=1.5$ surface-brightness
profile with small effective radii of $r_e=0.9$\,kpc, but  these
fake sources also do not correctly  represent the typical extended
Ly$\alpha$ surface-brightness profiles.  As the source detection
algorithms used in these surveys utilise parameters optimised for the
detection of compact sources, we argue that the inferred selection
functions in these studies must be too optimistic.  As we 
elaborate later, this leads to a bias in the luminosity function
estimate near the completeness limit of the surveys, thus leading to
incorrect estimates at the faint end of the LAE LF.  Moreover,  the
faint-end studies at $z\sim 6.5$ appear to be in subtle disagreement
\citep[][]{Ouchi2010,Matthee2015}.  Interestingly, \cite{Matthee2015}
followed a different approach  to \cite{Ouchi2010} to estimate
their completeness by rescaling fluxes of other sources in the
NB filter that do not show an excess but otherwise fulfil the
additional colour-selection criteria.  Nevertheless, this
model-independent approach, also utilised  in \cite{Sobral2017}, 
neglects that a significant fraction of Ly$\alpha$ emission comes from
the diffuse low-SB halo.

We argue here that assuming LAEs to be  compact point-like
sources is no longer a justifiable simplification.  As already
mentioned in Sect.~\ref{sec:non-param-results}, \cite{Grove2009}
suspected an inherent bias in LAE LF estimates caused by ignoring
possible extended emission in the construction of the selection
function.  Moreover, the LAEs found in the deep long-slit
  integration of \cite{Rauch2008}, and those  in the stacking analyses
by \cite{Steidel2011} and \cite{Momose2014}, already hinted at a large fraction
of LAEs being surrounded by low surface brightness Ly$\alpha$ halos.
Now, from the MUSE deep fields, the omnipresence of Ly$\alpha$ halos
around LAEs is a well-established fact on an object-by-object basis
\citep{Wisotzki2015,Leclercq2017}.  Here we show that accounting for
this effect results in an upward correction by a factor of up to three
for LF bins at
$\log L_\mathrm{Ly\alpha} [\mathrm{erg}\,s^{-1}] \lesssim 42.5$ of
previous surveys.

\section{Summary and outlook}
\label{sec:discussion}

We presented a framework for constructing the LAE LF in an integral
field spectroscopic survey.  We utilised these methods on the LAE
sample resulting from the first instalment of the MW survey.  Our LAE
LF sample covers luminosities
$42.2 \leq \log L_\mathrm{Ly\alpha} [\mathrm{erg\,s}^{-1}] \leq 43.5$.  We
show that the apparent LAE LF in this luminosity range is
non-evolving over the redshift range $2.9 \leq z \leq 6.7$.  This
result is irrespective of the assumed selection function, but we
argued that the classical assumption of LAEs being compact-point like
objects biases LF estimates too low near the completeness limit of a
survey.  We found that different non-parametric estimates provide
nearly identical descriptions of the cumulative or differential LAE
LF.  We obtained a maximum-likelihood Schechter parameterisation of
the LAE LF for
$\log L^* [\mathrm{erg\,s}^{-1}] = 42.66^{+0.22}_{-0.16}$, and
$\alpha = -1.84^{+0.42}_{-0.42}$, but with a strong degeneracy between
the two parameters.  The a posteriori normalisation of the
maximum-likelihood Schechter fit is
$\log \phi^* [\mathrm{Mpc^{-3}}] = -2.71$.  We show that the
Schechter parameterisation accurately describes our non-parametric
cumulative and differential estimates, while parametrising the LAE LF
with a simple power law provides a less optimal fit.  A comparison of
our LAE LF with binned estimates of the differential estimates from
the literature revealed subtle disagreements.  Especially at fainter
luminosities
($\log L_\mathrm{Ly\alpha} [\mathrm{erg\,s}^{-1}] \lesssim 42.5$), our LF
and the \cite{Drake2017a} MUSE HUDF LAE LF are higher than the
literature LF estimates.  This is a natural consequence of
incorporating the dilution of detectable Ly$\alpha$ signal due to
extended low surface brightness Ly$\alpha$ halos into the
completeness correction.  We show that we achieve a better agreement
with the literature when assuming for the completeness correction that
LAEs are compact point-like sources. However, in light of the recently
accumulated evidence regarding the ubiquity of extended Ly$\alpha$ halos, we
argued that this is an oversimplified assumption.

With the release of the full MW dataset (Urrutia et al., submitted to A\&A) we
will significantly improve the statistical robustness of the results 
presented here  by a factor of more than five,
due to the increased sample size.  The main drawbacks of the current data are the
lack of a sizeable sample of $z>5$ LAEs and the small number of very
luminous ($\log L_\mathrm{Ly\alpha} [\mathrm{erg\,s}^{-1}] > 43.0 $)
LAEs.  Even so,  it is especially this currently undersampled region in
the $(L_\mathrm{Ly\alpha},z)$-parameter space where other campaigns
hint at a possible evolution in the shape and normalisation of the LAE
LF \cite[e.g.][]{Santos2016,Sobral2017}.  While a robust determination
of the bright end of the LAE LF will only be possible within wide-area NB
campaigns, MW nicely populates the Ly$\alpha$ luminosity range that
overlaps with the faintest ends of such campaigns and the bright ends
of the MUSE deep surveys.  The next step in our analysis will be the
construction of a combined LAE LF from the final MW dataset and the
MUSE deep fields.

Of course, with an increased sample size on the horizon, we need to be
aware of possible systematic uncertainties in the framework  presented
here.  Firstly, all the  non-parametric and
parametric LF estimators applied here do not take photometric uncertainties into
account.  Secondly, we do not account for uncertainties in the
selection function.

Regarding the selection function construction we assumed that the ten
LAEs from the source insertion and recovery experiment in the HDFS are
representative of the whole population, and thus we weighted them
equally.  We can justify this approach, as no scaling relations
between Ly$\alpha$ halo flux fraction and other physical properties
have been found.  In particular, the halo-flux fraction appears to be
independent of Ly$\alpha$ luminosity \citep[][]{Leclercq2017}.  And,
as we explained in Sect.~\ref{sec:select-funct-from-2}, the 
sources used span a range of halo flux fractions and line profiles.
Nevertheless, to date  we do not have a 
for selection effects corrected distribution of halo flux fractions.  Equipped with such a
distribution in the future, a more realistic weighting scheme could be
employed.

However, a more relevant systematic effect might result from ignoring
the statistical errors on the flux measurement in the LF construction.
It is known,  especially near the completeness limit of a survey
where  the photometric uncertainties become larger, that ignoring
photometric errors systematically biases the LF.  This bias is
referred to as the \emph{Eddington-Malmquist} bias in the literature
\citep[see e.g. Sect. 5.5 in][]{Ivezic2014}.  The bias is a combined
effect of photometric errors, sample truncation on observed values,
and a rising luminosity function towards fainter luminosities.  The
effect is that near the completeness limit there are more sources that scatter into
the sample than sources that scatter out of the sample.  Ultimately
this results in higher inferred number source densities at the faint
end of the probed luminosity range, and thus  also biases the
inferred slopes steeper in parametric LF determinations.  We point out that
our sample truncation was quite conservative
(Sect.~\ref{sec:binn-trunc-sample}), i.e. we excluded almost 1/4 of
the faintest sources from our final LAE LF sample.  Moreover, in the
binned estimates the bin-size was chosen to be larger than the
photometric error in the faintest bin, and the  sources
scattering between the two faintest appear to compensate each other
in both directions.  A more quantitative discussion is beyond the
scope of this analysis, but we note that the
\emph{Eddington-Malmquist} bias has not been commented upon in the LAE
LF literature.  We argue that robust determinations of the faint-end
slope need to account for this bias in the future, for example  by modelling
the dependence of the photometric uncertainties on the inferred LFs.
Of interest in this respect appears the modified ML estimator
developed by \cite{Mehta2015} that can account for photometric
uncertainties.  Methods like this will allow for a robust and unbiased
determination of LAE LFs in the future, from which  vital 
information regarding cosmology and galaxy formation can in turn be extracted
(see e.g. \citealt{Bouwens2016} and \citealt{Dayal2018} for recent
reviews).


\begin{acknowledgements}
  We thank the support staff at ESOs VLT for help with the visitor
  mode observations during GTO.  This research made extensive use of
  the \texttt{astropy} pacakge \citep{AstropyCollaboration2013}. All
  plots in this paper were created using \texttt{matplotlib}
  \citep{Hunter2007}. E.C.H, R.S., T.U., and L.W. acknowledge funding
  from the Competitive Fund of the Leibniz Association through grants
  SAW-2013-AIP-4 and SAW-2015-AIP-2.  E.C.H. dedicates this paper to
  Melli's kittens Mary and Jamie.
\end{acknowledgements}

\bibliographystyle{aa}
\bibliography{bibliography.bib}

\end{document}